# Dependability in Embedded Systems: A Survey of Fault Tolerance Methods and Software-Based Mitigation Techniques

MOHAMMADREZA AMEL SOLOUKI, Politecnico di Torino, Italy
SHAAHIN ANGIZI*, New Jersey Institute of Technology, USA
MASSIMO VIOLANTE, Politecnico di Torino, Italy

Fault tolerance is a critical aspect of modern computing systems, ensuring correct functionality in the presence of faults. This paper presents a comprehensive survey of fault tolerance methods and software-based mitigation techniques in embedded systems. The focus is on real-time embedded systems, considering their resource constraints and the increasing interconnectivity of computing systems in commercial and industrial applications. The survey covers various fault-tolerance methods, including hardware, software, and hybrid redundancy. Particular emphasis is given to software faults, acknowledging their significance as a leading cause of system failures. Moreover, the paper explores the challenges posed by soft errors in modern computing systems. The survey concludes by emphasizing the need for continued research and development in fault-tolerance methods, specifically in the context of real-time embedded systems, and highlights the potential for extending fault-tolerance approaches to diverse computing environments.

## 1 INTRODUCTION

A key facet of fault tolerance is to ensure the continued correct operation of modern computing systems despite internal faults. The primary objective underlying fault tolerance endeavors is to increase system dependability. In a fault-tolerant system, the aim is to facilitate seamless transitions to alternative modules and thereby sustain service provision in the face of faults, by either concealing faults or detecting errors. To fulfill this aim, fault-tolerant systems must uphold specified service delivery, even amidst component faults [37].

---

*Corresponding author.

Authors' addresses: Mohammadreza Amel Solouki, mohammadreza.amelsolouki@polito.it, Politecnico di Torino, P.O. Box 10129, Turin, Italy; Shaahin Angizi, shaahin.angizi@njit.edu, New Jersey Institute of Technology, P.O. Box 07102, NJ, Newark, USA; Massimo Violante, massimo.violante@polito.it, Politecnico di Torino, P.O. Box 10129, Turin, Italy.

Failures arise when the behavior of a running system diverges from the system's expected behavior. Failures are caused by errors, while faults are the underlying cause of errors. Yet, it is noteworthy that not all faults necessarily lead to errors, and a single fault can precipitate multiple errors. Similarly, a solitary error can culminate in multiple failures. Redundancy, in some form, is an essential component across all fault tolerance approaches to ensure the system's capacity to withstand faults. Redundant devices, networks, data, or applications are leveraged based on the fault class at hand.

As of now, novel technologies elevate various facets of our quality of life while concurrently bolstering societal productivity and efficiency. Illustrative instances include innovative environmentally conscious transportation systems and advanced production methodologies, streamlining human effort and optimizing the generation of appliances and services. In the realm of automotive systems, the trajectory of emerging technological trends accentuates the introduction of novel features, expanding the array of onboard embedded systems and processors [39]. Within the automotive domain, these systems are engineered to optimize energy consumption, enrich user experiences through infotainment support, and institute autonomous and semi-autonomous control mechanisms encompassing methods like cruise control and autonomous piloting [101]. Furthermore, within the production sphere, burgeoning automation trends foster collaborative work environments uniting human workers and autonomous robots, thus amplifying production. This automation paradigm additionally seeks to mitigate human risk in scenarios involving hazardous conditions.

Both automotive and industrial production domains represent paradigmatic instances of safety-critical applications, wherein any functional malfunction of the supporting equipment, machinery, or devices could trigger dire repercussions, spanning critical injuries, fatalities, substantial property damage, or extensive environmental harm [58]. Consequently, the intricate electronic devices now integrated within these systems must rigorously adhere to safety, reliability, and security imperatives to ensure the seamless operation of the entire system.

Within the automotive domain, prominent corporations have invested, and are poised to continue investing, substantial capital in new technologies to not only implement but also broaden their applicability across various automotive functions. These applications encompass the development of diverse levels of vehicular autonomy driven by the attendant benefits to user safety, security, traffic latency reduction, and energy efficiency. Nevertheless, these technological advantages concurrently present various challenges yet to be definitively resolved. In principle, well-established methodologies for designing and developing secure and safe devices could be repurposed for use in these novel applications. However, both the automotive and autonomous machinery domains presently exploit a medley of innovative technologies, including Artificial Intelligence (AI) and computer vision, furnishing a distinct advantage in effecting more streamlined procedures. It's worth noting, though, that this trend equally introduces the ability for contemporary devices to integrate intricate algorithms, thereby augmenting application complexity and imposing substantial constraints concerning real-time operation, available power resources, and performance thresholds [39, 57, 101].

In practice, the development of modern safety-critical applications hinges upon three core elements: *i)* robust high-performance operation and power efficiency, *ii)* cost-effectiveness, and *iii)* unwavering safety and reliability [11]. In numerous instances, manufacturers and designers confront these demands by harnessing the latest transistor technology and scaling methods, thereby pushing the boundaries of Moore's law to incorporate an elevated transistor count within the same device. This endeavor yields appreciable enhancements in execution performance, power consumption, and practical production expenses.

However, various studies [7, 48, 50, 55, 79] have demonstrated that devices constructed using these cutting-edge technologies are inherently susceptible to an array of faults manifesting during initial operational stages and, with greater frequency, throughout their active lifespan. These faults may arise from two primary sources: *(i)* inherent defects stemming from manufacturing processes or component fatigue, and *(ii)* environmental or external influences [78]. In the former case, device faults might emanate from manufacturing anomalies that evade detection during end-of-production testing, thereby precipitating unforeseen behaviors during operational life-cycles. Furthermore, components within a device are predisposed to degradation (e.g., electro-migration or gate-oxide effects) following prolonged operation or even during periods of idleness (e.g., idle operational mode) [54], thereby potentially generating intermittent or permanent faults. In such scenarios, the faults arise due to aging or wear-and-tear effects [26, 42, 88]. Conversely, external influences also exert sway over device operation. Environmental factors temporarily or permanently alter electrical parameters, resulting in transient fault effects that impinge upon ongoing device applications. These fault effects propagate across the device as soft errors, which solely emerge when applications are executing on afflicted devices. Exposure to high-energy particles (triggering radiation effects) or electromagnetic interference (EMI) increases device vulnerability to transient faults, disrupting the electronic charge of one or more storage components within the device and toggling the state of transistors employed for data storage. As this data courses through the circuitry, multiple errors can arise within the application. In the most extreme instances, external interventions can lead to permanent damage to the device.

The fault-tolerance methods focus on detecting and recovering from faults, regardless of their types, to ensure the correct functioning of the system. To achieve a given reliability target, one commonly used fault-tolerance technique is the utilization of redundancy, in terms of hardware, software, information, and time, exceeding what is normally required for system operation. Hardware redundancy techniques involve adding extra hardware components to detect or tolerate faults. For example, multiple cores or processors can be utilized instead of a single one, with each application being executed on a separate core/processor, enabling fault detection and even correction. Another technique, time redundancy, allocates extra time to perform system functions and detect faults, without violating the timing constraints of real-time systems. The re-execution technique is an example of time redundancy, where a faulty task is repetitively executed on the same hardware until the correct output is obtained. Information redundancy techniques, such as error detection and correction coding, are commonly used in memory units, storage devices, and data communication to ensure reliability. Redundant Arrays of Independent Disks (RAIDs) are another example of information redundancy, where data is organized and stored in multiple configurations to enhance reliability. Additionally, software redundancy involves adding extra software to detect and tolerate faults. For example, N-version programming involves separate groups of programmers designing and coding a software module multiple times, reducing the likelihood of the same mistake occurring in all versions. Checkpointing, on the other hand, stores the last fault-free state of a process in stable memory, allowing the system to roll back to that state and re-execute the application in case of a fault. By employing these fault-tolerance methods, systems can ensure reliable functioning despite the occurrence of faults.

In this survey, we focus on fault-tolerance methods specifically tailored for embedded systems, considering their resource constraints, such as limited memory and low-end computation environments. However, our discussion covers techniques that can be adapted with minimal modifications to any general embedded system. Moreover, we believe that exploring the distinctive hardware-software constraints of resource-constrained embedded systems and leveraging real-time execution characteristics can lead to the development of fault-tolerance approaches applicable not only to embedded systems but also to real-time systems without resource limitations.

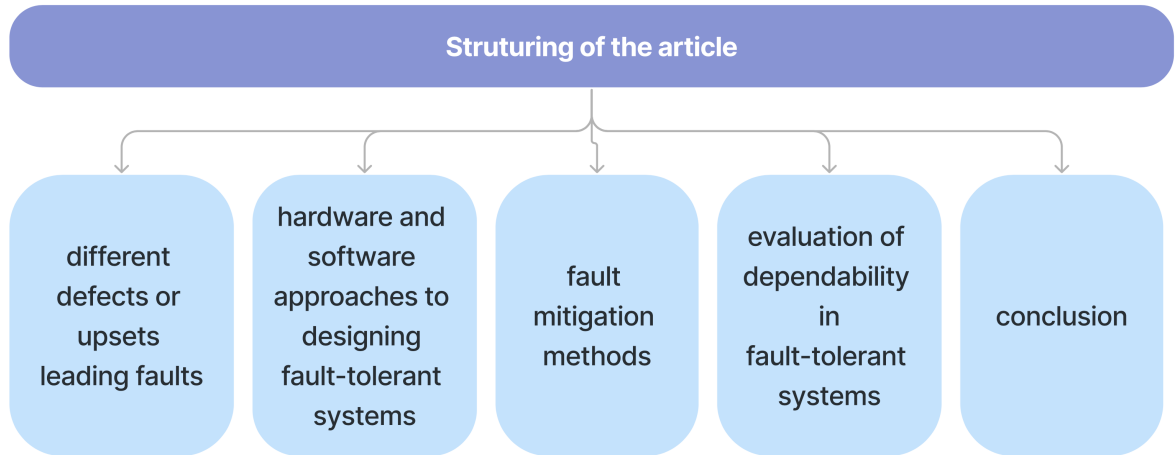

Fig. 1. Structuring of the article

The organization of this paper, illustrated in Figure 1, , is as follows. The background is reviewed in Section2. Section3 explores various hardware and software approaches to designing fault-tolerant systems. Section4 explores various fault mitigation methods. Section6 explores the evaluation of dependability in fault-tolerant systems. Finally, the conclusion is provided in Section7.

## 2 BACKGROUND

This section provides an overview of the different defects or upsets leading to permanent, intermittent, or transient faults. Specifically, the single event effects (SEEs) are discussed with a focus on Single Event Upsets (SEUs). Figure 2 visually represents the content discussed in this section.

Faults are the abstraction of a physical defect or upset at the logical level. In other words, faults describe the changes in device logic function caused by a defect or upset. Therefore, faults are defined here as any variation from the expected logical behavior of the underlying hardware. Faults can be further categorized as transient, intermittent, or permanent. Transient faults occur and then soon disappear. They manifest effects that can occur for a short period during the component's lifetime. Intermittent faults are characterized as a fault occurring, then vanishing, and then reoccurring, and so on. Examples of intermittent faults are signal interference, such as cross-talk between connections or communication lines. Permanent faults manifest and exist within the system until the defective component is repaired or replaced. These faults commonly occur due to manufacturing defects or physical damage to CMOS gates due to high charges. It is also possible that the device's electrical properties may mask some defects or upsets, causing no faults to appear.

Radiation effects can also lead to defects or upsets. Radiation exposure can result in both defects (permanent flaws) and upsets (temporary disturbances) in a system. Radiation is one of the external influences that can affect the behavior and reliability of electronic components and systems, potentially causing various types of faults.

### 2.1 Introduction to Single Event Effects

One of the most important radiation effects is Single-Event Effects (SEEs), which are a result of the interaction between an energetic particle and a semiconductor device, leading to various manifestations. SEEs are typically caused by the deposition of charge by the particle or by the creation of a current pulse. The amount of charge or current required to cause a SEE depends on the device's type and the materials used in its construction [81].

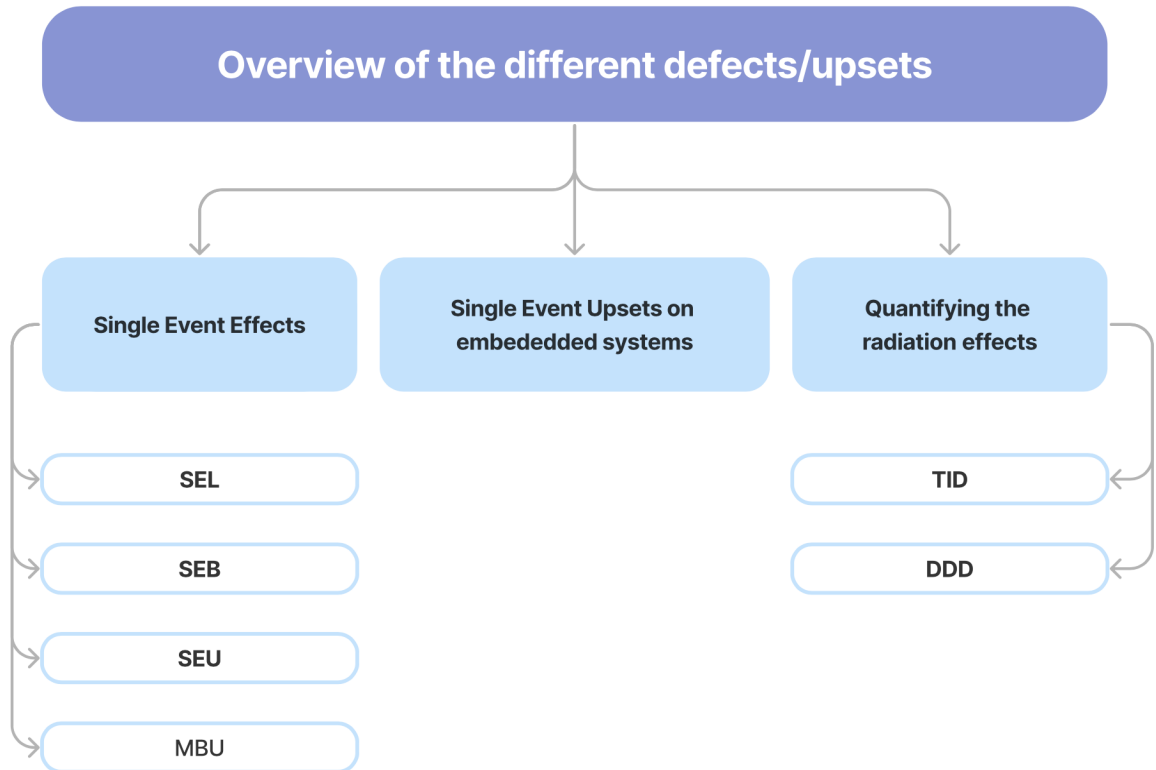

Fig. 2. Overview of the different defects/upsets.

When discussing SEEs, it is important to differentiate between permanent and transient effects [40]. Transient SEEs are temporary alterations in a device's state brought about by particle passage. These changes can stem from charge deposition or the creation of a current pulse. Transient effects refer to radiation-induced interference that ceases once the radiation dissipates. This temporary interference can involve variations in electrical signals, electronic device hardening, or system upsets that affect performance without causing permanent damage. Transient SEEs typically vanish within a few milliseconds and have no lasting impact on the device. One of the transient SEEs is the Single Event Transient (SET), which is a temporary change in the state of a device caused by the passage of a single energetic particle. SETs typically disappear after a few milliseconds, but they can cause errors in data processing.

On the other hand, permanent SEEs refer to changes in the device state that are caused by a particle's passage, resulting in irreversible damage that hampers proper functioning. Such effects encompass alterations in the semiconductor material's physical properties or degradation of circuitry, leading to long-lasting malfunctions or failures. Examples of permanent SEEs include Single-Event Latch-up (SELs), Single-Event Burnouts (SEBs), Single-Event Upsets (SEUs), and Multi-Bit Upsets (MBUs) [53].

SEL is a type of SEE where a high-current path is formed in the device, leading to permanent damage. SEL occurs when a transistor turns on and stays on, even when the gate voltage is removed. SEB is a type of SEE by a single energetic particle passage, like a high-energy ion or neutron, resulting in irreversible damage. This phenomenon occurs when the particle deposits excessive energy, leading to localized heating and damage within the device's structure. This can result in a sudden increase in current or voltage, causing permanent damage or burnout. SEU is a permanent change in the state of a memory cell or register caused by the deposition of energy in the semiconductor material by a single energetic particle. It can cause a single bit to be flipped from a 0 to a 1 or vice versa [20, 112]. MBU refers to the SEU of multiple memory cells in close proximity caused by a single energetic particle. MBUs are less common than SEUs, but they can be more serious.

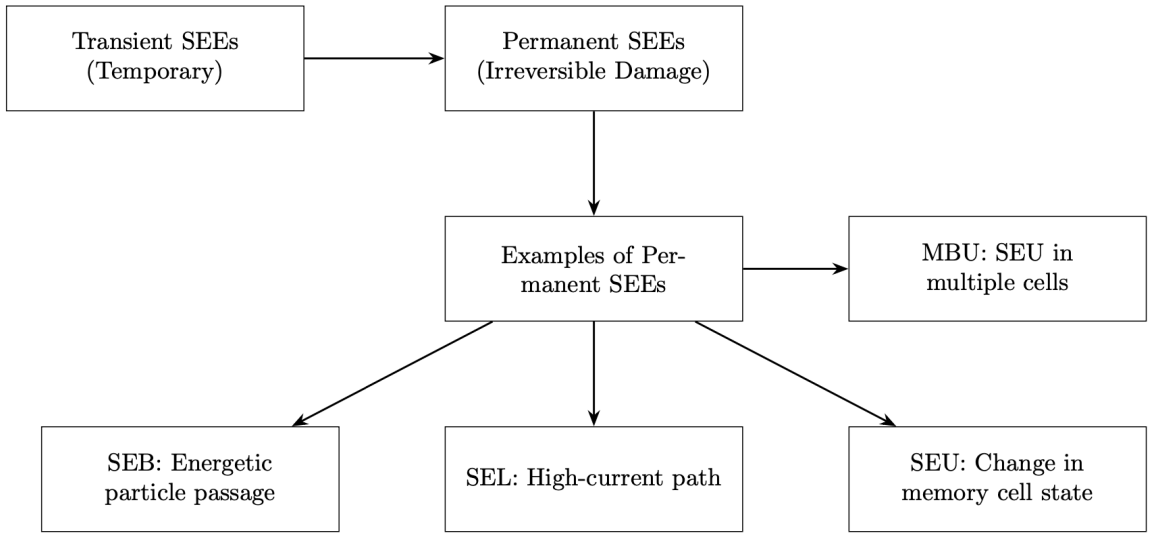

Fig. 3. Classification of Single-Event Effects (SEEs)

To conclude, SEEs are significant radiation-induced phenomena resulting from particle interactions with semiconductor devices. They can be categorized into transient effects, which are temporary disruptions with no lasting impact, and permanent effects, causing irreversible damage to devices. Examples of permanent SEEs include SELs , SEBs , SEUs , and MBUs, each with specific consequences for device functionality. Figure 3 shows the classification chart, providing a visual representation that enhances the comprehension of SEEs.

## 2.2 The Impact of Single Event Upsets on Embedded Systems

This subsection delves into the intricate details of SEUs, encompassing their diverse impacts on various components of embedded systems. In the realm of fault tolerance, it's crucial to distinguish between Silent Data Corruption (SDC) and Single Event Functional Interrupt (SEFI). SDC encompasses errors in memory and the final application output, while SEFI, a severe issue causing system hangs or crashes, directly impacts application execution and user experience. Most fault tolerance methods primarily target SDC, leaving SEFI, which can significantly disrupt system operation, relatively unaddressed [59, 109].

Figure 4 in the context of SEUs, it provides a visual representation of how these events affect different components.

However, it's imperative to elaborate on SDC and SEFI for clarity. SDC occurs when memory or final application output is corrupted, leading to inconsistencies in results. In contrast, SEFI, resulting from errors in control flow, leads to application crashes and processor hang. Only a few techniques can effectively detect both SDC and SEFI, such as those employing the lockstep principle based on redundancy to enhance processor dependability. SEUs can affect data flow or control flow in processors, influencing data flow errors or SEFIs [105].

Moving forward, this subsection primarily focuses on SEUs in detail. In the Register File, SEUs can corrupt data and cause errors in the application outputs, leading to inconsistencies in the results. If an SEU impacts a control register, it can result in errors in the execution flow of the program, and the system freezes. SEUs in the Integer Unit (IU) and Floating Point Unit (FPU) can lead to incorrect computations due to the pipelining in these arithmetic units. In the Bus Unit, bit flips in the embedded registers responsible for latching addresses and data can cause incorrect read or write operations. The Control Unit, which implements complex algorithms, may experience SEUs that trigger exception generation or disrupt the sequence. SEUs can also affect the Debug Unit, activating special execution modes and causing errors in the program's execution flow.

Moving on to the Instruction Cache, SEUs can result in corrupted outputs or processor freezes. The instruction caches typically consist of an SRAM array for storing fetched instructions and a tag array for validating or invalidating the fetched program. SEUs in the tag array can invalidate an instruction to be executed, leading to a cache miss and introducing a delay in program execution as the instruction needs to be fetched again. If an SEU validates an incorrect code, it can crash the program's flow. Additionally, an SEU can corrupt an instruction in the SRAM array. If the tag array validates this corrupted code, a wrong instruction will be executed, or an exception will be generated if the corrupted instruction is no longer part of the processor instruction set. However, if the tag array does not validate the corrupted instruction, the fault is masked, and no incorrect behavior is observed. The Instruction Cache section in Figure 4 captures these dynamics. Similar to instruction caches, Data Caches also consist of a tag array and a data array. Bit flips in the tag array can validate outdated data, resulting in incorrect outputs or invalidate data, causing delays (cache miss) in the application. If an SEU affects the data array, it can corrupt the output. However, if the data is outdated, the fault is masked, and no effects are observed [52].

## 2.3 Quantifying the Radiation Effects

To quantify the effects of radiation on electronic devices, various measurements have been developed. This article will discuss the most commonly used measures: Total Ionizing Dose (TID), and Displacement Damage Dose (DDD) [50, 52]. Total Ionizing Dose (TID) is the total amount of ionizing radiation absorbed by an electronic device over time. This radiation can lead to permanent damage, such as changes in the electrical properties of the semiconductor material, degradation of the circuitry, or an increase in leakage currents. TID is measured in units of rads (radiation absorbed dose) or grays (Gy). The amount of TID that a device can withstand before failing depends on the type of device and the materials used in its construction. TID can cause permanent damage to a device by creating defects in the semiconductor material. These defects can reduce the conductivity of the material or create hot spots that can lead to thermal breakdown. DDD is the amount of non-ionizing radiation that causes permanent damage to the crystal lattice of the semiconductor material. It can lead to changes in material properties, which can affect the performance of electronic devices. DDD is measured in units of displacements per atom (dpa).

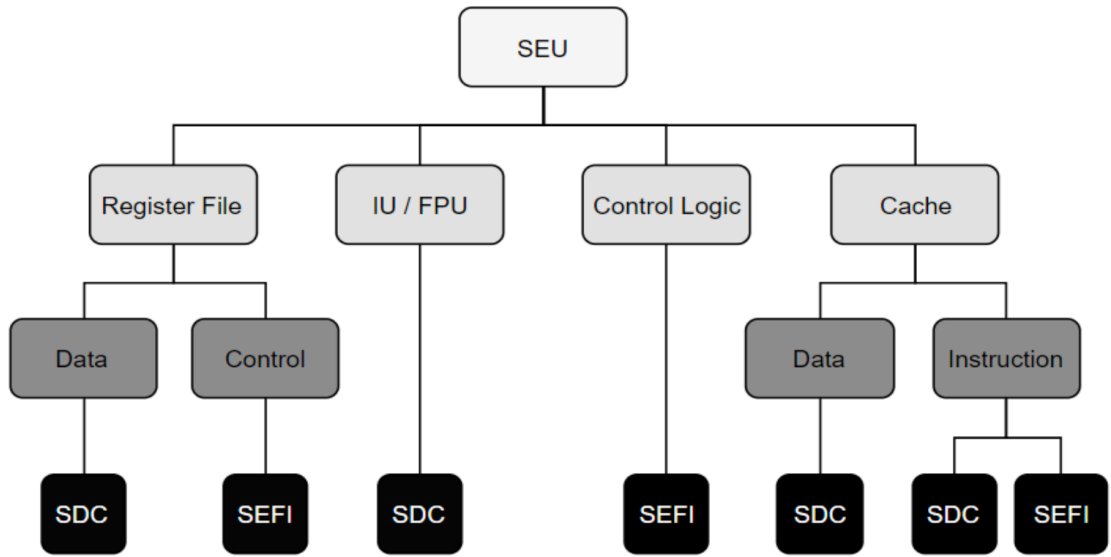

Fig. 4. The impact of single error upsets on different parts of a processor [77].

The amount of DDD that a device can withstand before failing depends on the type of device and the materials used in its construction. DDD can cause permanent damage to a device by creating defects in the semiconductor material. These defects can reduce the conductivity of the material or create hot spots that can lead to thermal breakdown.

## 2.4 Fault Impact on Program Execution

Understanding the impacts of faults on program execution is crucial for designing fault-tolerant embedded systems. Faults in such systems can lead to undesired outcomes, including program crashes, incorrect outputs, and compromised system functionality. This subsection provides an overview of the effects of faults on program execution, emphasizing the significance of understanding these effects in the context of fault mitigation methods.

- Program Crashes and Abnormal Terminations: One of the primary consequences of faults in program execution is program crashes and abnormal terminations. Faults such as hardware failures, memory corruption, or unhandled exceptions can cause the program to terminate abruptly or enter an undefined state, resulting in system instability and potential data loss [37]. Researchers have proposed various techniques for detecting and recovering from program crashes, including Control Flow Checking methods that verify the integrity of program execution path
- Incorrect Outputs and Results: Faults can lead to incorrect outputs and results, affecting the reliability and accuracy of embedded systems. Logic errors, data corruption, or faulty computations can result in incorrect data processing and decision-making, leading to undesirable consequences [72]. To mitigate the impact of such faults, researchers have explored techniques such as redundant computation, error-correcting codes, and diverse redundancy approaches to ensure accurate and reliable output generation.
- Performance Degradation: Faults in embedded systems can also cause performance degradation, resulting in decreased system efficiency and responsiveness. Resource management errors, such as memory leaks and inefficient scheduling, can lead to performance bottlenecks [41]. Researchers have investigated methods such as dynamic resource allocation, optimized scheduling algorithms, and memory management techniques to mitigate performance degradation caused by faults.
- Security Vulnerabilities: Faults in embedded systems can introduce security vulnerabilities, jeopardizing the confidentiality, integrity, and availability of sensitive data. Faults such as input validation flaws, buffer overflows, or insecure communication protocols can be exploited by attackers to gain unauthorized access or perform malicious activities [90]. Researchers have proposed security-oriented fault mitigation methods, including secure coding practices, encryption algorithms, and intrusion detection systems.
- Data Corruption: Faults in embedded systems can lead to data corruption, compromising the reliability and integrity of stored data. Faults such as power failures, communication errors, or hardware malfunctions can result in data inconsistencies or loss [60]. To mitigate data corruption, researchers have explored techniques such as checksums, error detection and correction codes, and redundant storage mechanisms.

In summary, understanding the effects of faults on program execution is essential for designing fault-tolerant embedded systems. By considering the potential consequences of faults, researchers can develop effective fault mitigation methods.

## 3 DESIGNING FAULT-TOLERANT SYSTEMS: HARDWARE AND SOFTWARE APPROACHES

This section aims to investigate a range of hardware and software techniques for designing fault-tolerant systems. subsection 3.1 concentrates on Hardware-based fault tolerance methods. subsection 3.2 extensively discusses Software-based Fault Tolerance methods, specifically the Single-Design Software Fault Tolerance methods. Lastly, subsection 3.3 presents Hybrid methods that incorporate a blend of hardware and software methods.

The choice of fault tolerance approach to be employed is contingent upon the specific application in question. For instance, a system that requires high availability, such as a telecommunications network, would typically opt for hardware-based fault tolerance methods. Conversely, a less critical system like a word processing application may utilize software-based fault tolerance methods [37, 61].

The organization of this section, as depicted in Figure 5, is as follows, providing a visual representation of each section.

### 3.1 Hardware based Fault Tolerance Methods

Hardware-based techniques in fault detection and correction can be categorized into two main groups: hardware monitors and redundancy-based. Hardware monitor methods are employed to detect faults, while redundancy-based methods are employed for fault mitigation.

*3.1.1 Hardware Monitors.* Hardware monitors are specialized components or circuits integrated into a system to monitor the behavior of various components and signals continuously. These monitors are designed to detect anomalies, errors, or deviations from expected behavior, which could indicate the presence of faults or defects in the system. By actively monitoring the system's operation in real-time, hardware monitors can provide early warnings and trigger appropriate actions to prevent or mitigate the effects of faults before they lead to system failures.

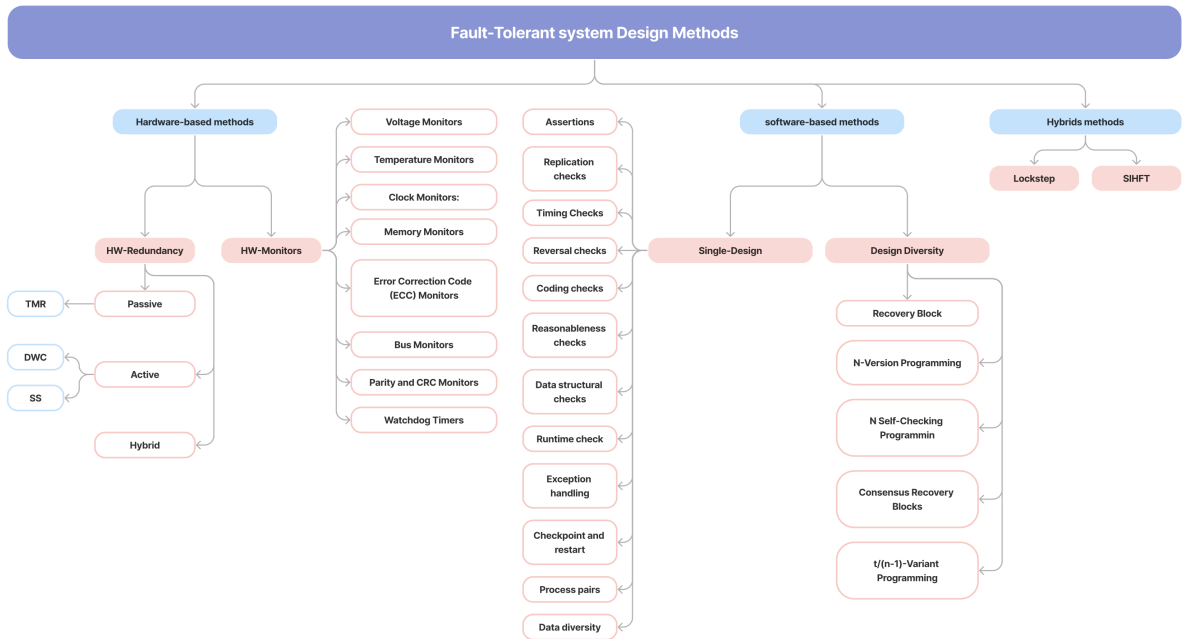

Fig. 5. Fault-tolerant system design methods

There are several types of hardware monitors, each serving a specific purpose:

- Voltage Monitors: These monitors supervise the supply voltage levels of critical components. If the voltage falls outside specified limits, it might indicate a fault or power-related issue.
- Temperature Monitors: Monitoring the temperature of components is essential to prevent overheating and thermal damage. Temperature sensors and monitoring circuits can trigger alerts or take actions to cool down the system if temperatures become excessive.
- Clock Monitors: These monitors oversee clock signals to ensure proper timing and synchronization between components. Deviations in clock frequencies or signal integrity can lead to faults.
- Memory Monitors: Monitoring memory operations helps identify errors in data storage or retrieval, which is crucial for maintaining data integrity.
- Error Correction Code (ECC) Monitors: ECC monitors detect and correct errors in memory or data storage systems using error-correcting codes, thereby enhancing data reliability.
- Bus Monitors: These monitors supervise data and control buses for communication errors between components. Detecting bus errors can prevent data corruption or incorrect communication.
- Parity and CRC Monitors: These monitors use parity or cyclic redundancy check (CRC) codes to detect data corruption during transmission or storage.
- Watchdog Timers: Watchdog timers are hardware-based timers that must be periodically reset by the system's software. If the software fails to reset the timer within a specified time frame, the watchdog timer assumes a fault has occurred and initiates a system reset.

Hardware monitors work in conjunction with redundancy-based techniques to provide comprehensive fault detection and tolerance mechanisms. They contribute to creating fault-tolerant systems that can identify, isolate, and recover from faults, ultimately enhancing the overall reliability and availability of critical electronic systems.

*3.1.2 Hardware Redundancy.* Redundancy-based techniques rely on hardware or time redundancy. These techniques involve the addition of extra hardware components to detect or tolerate faults. such as watchdog processors [68], checkers [14], or infrastructure intellectual properties (I-IP) [66].

Hardware redundancy can be implemented through passive, active, and hybrid methods.

Passive redundancy techniques use M-of-N systems where N components are present, and correct system operation is achieved when at least M components work correctly. For instance, Triple Modular Redundancy (TMR) is a 2-of-3 system, meaning it consists of three components performing the same action, and the result is voted on to determine the correct output [37, 62].

Active redundancy techniques include duplication with comparison (DWC), standby-sparing (SS), pair-and-a-spare, and watchdog timers. DWC involves parallel execution of two identical hardware components, with the output being compared to detect faults. However, DWC can only detect faults and not tolerate them. Standby-sparing utilizes one operational module and one or more spare modules. If a fault is detected in the main component, it is omitted from operation, and the spare component takes over. Pair-and-a-spare is a combination of DWC and SS techniques, where two modules are executed in parallel, and their results are compared to detect faults [37, 62].

Hybrid redundancy techniques integrate features from both active and passive hardware redundancies. Examples of hybrid redundancy include N modular redundancy with spare, sift-out modular redundancy, self-purging redundancy, and triple duplex architecture. Self-purging redundancy is based on NMR with spare techniques, where all modules actively participate in the system function. Sift-out modular redundancy utilizes special circuits, such as comparators, detectors, and collectors, to configure N identical modules in the system. Triple duplex architecture combines DWC with TMR to detect faulty modules and remove them from the system. These hardware-based techniques, although effective in fault detection and tolerance, come with a high cost, verification and testing time, area overhead, and increased power consumption [37, 62].

In summary, Hardware-based fault detection and correction techniques fall into two main categories: hardware monitors and redundancy-based methods. Hardware monitors continuously observe system components and signals, detecting anomalies to provide early warnings and prevent faults. These monitors include voltage, temperature, clock, memory, ECC, bus, and parity/CRC monitors, as well as watchdog timers. These work in tandem with redundancy-based techniques, which involve adding extra hardware components to detect or tolerate faults. Redundancy methods can be passive (M-of-N systems), active (DWC, SS, pair-and-a-spare), or hybrid, combining features of both. While effective, these techniques come with costs such as verification, testing, area overhead, and power consumption.

## 3.2 Software based Fault Tolerance Methods

Software based Fault Tolerance Methods can be divided into two categories: Design Diversity-Based and single-design software fault tolerance. In Design Diversity-Based methods, multiple diverse versions of a software module are created, often using different algorithms or programming languages. These versions run concurrently, and discrepancies are detected and resolved through voting mechanisms, enhancing reliability. On the other hand, single-design software fault tolerance focuses on enhancing the robustness of a single software design through techniques such as error detection, error handling, and recovery mechanisms.

*3.2.1 Design Diversity Based Software Fault Tolerances.* Design Diversity-Based or Multiple-Version-Based software fault tolerance involves using multiple versions or variants of software, either executed sequentially or in parallel. These versions are used as alternatives, with separate means of error detection, and can be implemented in pairs or larger groups for replication checks or masking through voting. The main idea is that components built differently should fail differently, so if one version fails on a specific input, at least one alternate version should be able to produce the correct output. This section explores various approaches to software reliability and safety through design diversity. However, ensuring the independence of failure among multiple versions and developing effective output selection algorithms are critical challenges in deploying multi-version software fault tolerance techniques.

Design diversity serves as a means of protection against uncertainty, specifically, design faults and their associated failure modes in software design. The objective of applying design diversity techniques to software design is to build program versions that fail independently and with a low probability of coincidental failures. Achieving this objective greatly reduces or eliminates the probability of encountering incorrect outputs during program execution. However, due to the complexity of software, the application of design diversity for software fault tolerance is currently more of an art than a science.

The concept of multiple-version software design was pioneered by Algirdas Avizienis and his team at UCLA in the 1970s, primarily focusing on software. Their research also explored the application of design diversity concepts to other system aspects such as the operating system, hardware, and user interfaces. Even with rigorous development and proper application of design diversity, there is still the issue of identical input profiles leading to common errors. Experiments have shown that error manifestations are not equally distributed across the input space, and the probability of coincident errors is influenced by the chosen inputs. Data diversity techniques can potentially mitigate this issue, but quantifying their effectiveness remains a challenge.

An important consideration in using multi-version software is the cost involved. Replicating the entire development effort, including testing, would be expensive. In some cases, where only certain parts of the functionality are safety-critical, applying design diversity only to those critical parts can reduce development and production costs. [95] highlights the need to address the problem of identical input profiles as a common source of errors, highlighting that experiments have indicated unequal distribution of error manifestations across the input space. While data diversity techniques may reduce the impact of this error source, quantifying their effectiveness remains a challenge.

In summary, Design Diversity-Based or Multiple-Version-Based software fault tolerance offers a means of enhancing software reliability and safety by using multiple versions of software with independent failure properties. However, challenges exist in ensuring independence from failure and developing suitable output selection algorithms. The concept of design diversity has evolved as an art in software fault tolerance, with applications extending beyond software to other system aspects. The issue of identical input profiles leading to common errors requires attention, and while data diversity techniques may mitigate this, quantifying their effectiveness remains a challenge. The cost of using multi-version software is an important consideration, and selectively applying design diversity to critical parts can help reduce development and production costs.

In this study, we explore various fault-tolerance approaches in software that incorporate design diversity, both with multiple versions and a single design. The approaches we focus on are as follows:

- The Recovery Block Scheme: The Recovery Block Scheme (RBS) combines the checkpoint and restart approach with multiple versions of a software component [89]. Before execution, checkpoints are created to allow for recovery after detecting errors. This ensures a valid operational starting point for the next version if an error is detected. Additionally, embedded checks are used to enhance error detection. The primary version executes more frequently compared to alternates, which are designed for degraded performance. Multiple versions can be executed sequentially or in parallel, depending on processing capability and desired performance. In the event that all alternates fail, the component must raise an exception to communicate its failure to the system.

- The N-Version Programming Scheme: The N-Version Programming Scheme (NVPS) is a multiple-version technique where all versions fulfill the same basic requirements, and the correctness of output decisions relies on comparing all outputs [15]. A voter selects the correct output, eliminating the need for an acceptance test based on the application. Developing NVPS requires considerable effort as all versions must adhere to the same conditions, resulting in complexity comparable to creating a single version. Designing the voter can be challenging and may involve inexact voting. Different voters, such as the Formalized Majority Voter, Generalized Median Voter, Formalized Plurality Voter, and Weighted Averaging Techniques, can be used, with weights based on the application and individual versions' features.
- The N Self-Checking Programming Scheme: The N Self-Checking Programming Scheme (NSCPS) combines various structural variations of Recovery Blocks and N-Version Programming using multiple software versions [63]. Independent development of versions and acceptance tests based on shared requirements are used in this technique. NSCPS utilizes separate acceptance tests for each version, distinguishing it from the Recovery Blocks approach. The technique benefits from using an application-independent decision algorithm for selecting the correct output.
- The Consensus Recovery Blocks Scheme: The Consensus Recovery Blocks Scheme (CRBS) combines N-Version Programming and Recovery Blocks to achieve higher reliability compared to either approach individually [97]. The acceptance test in Recovery Blocks techniques lacks guidance and may have design faults, whereas voters in N-Version Programming can be unsuitable in certain cases. CRBS incorporates the first layer of decision-making using a similar algorithm to that of N-Version Programming. If the first layer declares a failure, the second layer, which utilizes acceptance tests similar to Recovery Blocks, is invoked. Although more complex than the individual techniques, CRBS has the potential to deliver a more reliable result.
- The t/(n-1)-Variant Programming Scheme: The t/(n-1)-Variant Programming Scheme (VPS) involves n variants and the t/(n-1) diagnosability measure to restrict faulty units to a subset of size at most (n-1), assuming a maximum of t faulty units. This approach differs from the previous methods in terms of the methodused to isolate faulty units [85].

In summary, the utilization of Design Diversity-Based or Multiple-Version-Based software fault tolerance techniques offers promising avenues to enhance software reliability and safety. These approaches leverage multiple versions of software, designed to fail independently, thereby reducing the likelihood of encountering erroneous outputs during program execution. However, the practical implementation of design diversity in software fault tolerance remains more of an art than a science due to the complexity of software and the challenges in ensuring independence from failure. Additionally, addressing the issue of identical input profiles leading to common errors and quantifying the effectiveness of data diversity techniques remain significant challenges. Furthermore, the cost implications of employing multi-version software must be carefully considered, and selective application of design diversity to critical components can help mitigate development and production expenses. The various fault-tolerance approaches explored, such as the Recovery Block Scheme, N-Version Programming Scheme, N Self-Checking Programming Scheme, Consensus Recovery Blocks Scheme, and t/(n-1)-Variant Programming Scheme, provide diverse strategies to implement design diversity effectively in software fault tolerance.

*3.2.2 Single-Design Software Fault Tolerance Approach.* Single-design fault tolerance is a method that involves introducing redundancy to a single version of the software in order to detect and recover from faults. In the context of single-version software fault tolerance techniques, various factors need to be considered, including program structure, error detection, exception handling, checkpoint and restart, process pairs, and data diversity.

In terms of software engineering aspects, the use of modularizing techniques is crucial for implementing fault tolerance effectively. Modular decomposition should include built-in protections to prevent abnormal behavior from propagating to other modules. Control hierarchy issues, such as visibility and connectivity, should also be taken into account to minimize the risk of uncontrolled corruption of the system state. Partitioning can provide isolation between functionally independent modules, leading to simplified testing, easier maintenance, and lower propagation of side effects. System closure, which states that no action is allowed unless explicitly authorized, is another important principle of fault tolerance. Atomic actions, which are activities in which components exclusively interact with each other without any interaction with the rest of the system, offer error confinement and recovery capabilities. If an atomic action terminates normally, its results are complete and committed. If a failure occurs during an atomic action, it only affects the participating components [113].

To ensure the effective application of fault tolerance techniques in single version systems, structural modules should possess two basic properties: self-protection and self-checking. Self-protection means that a component can detect errors in the information passed to it by other interacting components. Self-checking means that a component can detect internal errors and take appropriate actions to prevent error propagation. The extent to which error detection mechanisms are used in a design depends on the cost of additional redundancy and the run time overhead. It's important to note that fault tolerance redundancy is not intended to contribute to system functionality but rather to the quality of the product. Similarly, detection mechanisms can affect system performance. The utilization of fault tolerance in a design involves trade-offs between functionality, performance, complexity, and safety.

Assertions [92], which are logical statements inserted at different points in a program reflecting relationships between program variables, can also be used for fault tolerance. However, their effectiveness depends on the nature of the application and the programmer's ability. Control Flow Checking (CFC) involves partitioning the application program into basic blocks (BBs) and computing deterministic signatures for each block. Faults can be detected by comparing the run time signature with a precomputed one.

The authors [64] have proposed a classification of error detection checks, some of which can be selected for implementing the mentioned module properties. The checks can be located either within the modules or at their outputs, depending on the requirements. The checks encompass replication, timing, reversal, coding, reasonableness, and structural checks.

Replication checks involve matching components with error detection based on the comparison of their outputs, making them suitable for multi-version software fault tolerance [100]. Timing checks are applicable to systems and modules with timing constraints and can look for deviations from acceptable module behavior [37]. Watchdog timers, a type of timing check, can be used to monitor system behavior and detect "lost or locked out" components. Reversal checks use the output of a module to compute the corresponding inputs and detect errors if the computed inputs do not match the actual inputs. Coding checks utilize redundancy in the representation of information and check relationships between actual and redundant information before and after operations. Reasonableness checks rely on semantic properties of data, such as range, rate of change, and sequence, to detect errors. Data structural checks involve inspecting known properties of data structures, such as number of elements, links, and pointers. Augmenting data structures with redundant structural data can enhance the effectiveness of structural checks. Runtime checks are standard error detection mechanisms in hardware systems and can be used as fault detection tools [85]. Fault trees, top-down graphical representations of failures and triggering conditions can aid in the development of fault detection methods by identifying failure classes and triggering conditions.

Exception handling involves interrupting normal operations to handle abnormal responses. Exceptions are signaled by error detection mechanisms, and the design of exception handlers requires consideration of possible triggering events, their effects on the system, and appropriate recovery actions [85].

Checkpoint and restart is a common recovery method for single-design software. Most software faults that occur after development are unanticipated, state-dependent faults. Restarting a module is usually sufficient to complete its execution successfully. Restart recovery can be static or dynamic. Static restart returns the module to a predetermined state, while dynamic restart uses dynamically created checkpoints [85].

Process pairs utilize two identical versions of software running on separate processors. The recovery method is a checkpoint and restart. The primary processor actively processes input and creates output while generating checkpoint information for the backup processor. Upon error detection, the secondary processor loads the last checkpoint and takes over the primary processor's role. The faulty processor goes offline for diagnostic checks. This technique ensures uninterrupted delivery of services after a failure [85].

Data diversity is an effective defense method against design faults, especially when combined with checkpoint and restart methods. By implementing "input sequence workarounds" and using different input re-expressions on each retry, data diversity enhances the success rate of checkpoint and restart procedures. The desired outcome of each retry is to generate output results that are either exactly the same or semantically equivalent, although the definition of equivalence may vary depending on the application. In [73], three fundamental data diversity models are presented: (i) Input Data Re-Expression, which focuses on modifying the input; (ii) Input Re-Expression with Post-Execution Adjustment, which involves processing the output to achieve the desired value or format; and (iii) Re-Expression via Decomposition and Recombination, where the input is broken down into smaller elements and then recombined after processing to obtain the desired output. It is worth noting that data diversity works hand in hand with the Process Pairs technique, allowing for different re-expressions of the input in the primary and secondary.

In the context of operating systems, software fault tolerance is crucial to ensure the proper functioning of any application-level software. While designing and building operating systems can be complex, time-consuming, and costly, it may be necessary to develop custom operating systems with highly structured design processes involving experienced programmers and advanced verification techniques for safety-critical applications. Another approach to achieving fault tolerance in operating systems for mission-critical applications is to use wrappers on off-the-shelf operating systems to enhance their robustness against faults. However, utilizing off-the-shelf software on dependable systems poses the challenge of ensuring the reliability of the components for the intended application. It is known that the development process for commercial off-the-shelf software lacks consideration for safety or mission-critical standards, resulting in weak documentation for design and validation activities. On the other hand, commercial operating systems offer advantages such as incorporating the latest developments in operating system technology and potentially having fewer bugs overall due to continuous bug-fixing efforts driven by user complaints. In order to minimize the risk of introducing design faults, it is preferable to adopt techniques that utilize the operating system as is, without internal modifications. Wrappers serve as middleware between the operating system and application software, monitoring the flow of information to prevent undesirable values from propagating. By limiting the input and output spaces of a component, wrappers provide application-transparent fault tolerance functionality. In [94], wrappers referred to as "sentries" encapsulate operating system services and can modify the characteristics of these services as perceived by the application layer. Through wrappers, fault-tolerance methods can be dynamically assigned to specific applications based on their individual needs in terms of fault tolerance, cost, and performance. Authors proposed using wrappers at the micro-kernel level for off-the-shelf operating systems, aiming to verify semantic consistency constraints using abstractions or models of the expected component functionality.

In conclusion, Software based fault tolerance methods offer several advantages, including the absence of additional auxiliary devices, no specific operating system requirements, good expansibility, and support for continuous exploration and repeated experiments. However, these methods come with significant time and space overhead due to the inclusion of numerous redundant instructions, which can significantly impact program performance.

Table 1. Overview of the techniques classification. [77]

| Technique Classification | Pros | Cons |
| --- | --- | --- |
| **Hardware** | -High fault detection<br>-Fast detection<br>-No software modification | -Most does not correct errors<br>-Mainly single fault model<br>-High area and power overhead<br>-Implemented only in physical level<br>-Can be expensive |
| **Software** | -High fault detection<br>-High flexibility<br>-No hardware modification<br>-Small area overhead<br>-Some can correct errors | -High performance overhead<br>-Mainly single fault model<br>-Focuses only on data or control flow, but not both |
| **Hybrid** | -High fault detection<br>-High efficiency<br>-Can achieve small area overhead<br>-Some can detect both SDC and SEFI<br>-Some can correct errors | -Can also achieve high performance or area overhead<br>-Software and hardware modification |

## 3.3 Hybrid Methods

Hybrid fault-tolerance methods typically involve the integration of a Software Implemented Hardware Fault Tolerance (SIHFT) method with a hardware module designed to perform consistency checks within the processor. In a study by [33], SIHFT techniques are combined with a Control Flow Checking (CFC) module, which is responsible for monitoring the trace port of the processor. Another hybrid approach, proposed by [17], is known as Hybrid Error-detection Technique using Assertions (HETA). This method utilizes a watchdog module and assertions (or signatures) to address control-flow errors.

Lockstep is another hybrid fault-tolerance technique that utilizes both software and hardware redundancy for error detection and correction [5],[110],[45],[83]. Lockstep involves executing the same application simultaneously and symmetrically in two identical processors. These processors are initialized to the same state and receive identical inputs during system start-up. During normal operation, the state of both processors should be identical at each clock cycle. By monitoring the processor's data, addressing, and controlling buses [27], a checker module periodically compares the outputs of the processors to check for inconsistencies. To enforce verification, specific points are inserted in the program to indicate when the application execution should be locked and the outputs compared. If any discrepancies are found, the lockstep system leverages a rollback methodto restore the processors to a safe state. In the absence of errors, a checkpoint operation is performed, which stores the context of the processor (including registers and main memory) in a secure memory location. Memories can be protected using Error Correction Code (ECC) to prevent data corruption. ECC is capable of detecting and correcting single-bit errors and detecting double-bit errors. To recover from errors, the fault-free copy of the processor's context is retrieved from memory using the rollback method. The processor is then recovered to a state without errors and restarts the application execution from this point.

In summary, hybrid fault-tolerance methods combine software and hardware approaches to enhance error detection and correction. One approach integrates Software Implemented Hardware Fault Tolerance (SIHFT) with Control Flow Checking (CFC) or Hybrid Error-detection Technique using Assertions (HETA) to monitor and address control-flow errors. Another hybrid method, known as Lockstep, executes applications in parallel on identical processors, comparing outputs and employing rollback and checkpoint mechanisms to ensure system reliability and error recovery. These hybrid approaches provide robust fault tolerance in critical systems.

Table 1 provides a comprehensive classification of techniques for addressing random hardware failures (RHFs) in embedded systems. The techniques are categorized into three main types: hardware-based, software-based, and hybrid-based approaches. In the hardware category, these techniques offer advantages such as high fault detection rates, fast detection capabilities, and the absence of software modifications. However, they come with notable drawbacks, including the inability to correct detected errors, a predominant focus on a single fault model, substantial area and power overhead, implementation restricted to the physical level, and potentially high cost. Software-based techniques, on the other hand, boast high fault detection rates, flexibility, and minimal hardware modifications, with some capable of error correction. Nevertheless, they incur drawbacks such as high-performance overhead, a predominant focus on a single fault model, and concentration on either data or control flow, but not both. Hybrid techniques aim to combine the strengths of hardware and software approaches, achieving high fault detection efficiency, small area overhead, and the capability to detect and, in some cases, correct both SDC and SEFI. However, hybrid techniques also have their challenges, including the potential for high performance or area overhead and necessitating both software and hardware modifications. The insights provided by Table 1 pave the way for a nuanced understanding of the strengths and limitations associated with each category of techniques, facilitating informed decisions in selecting and implementing RHF mitigation strategies in embedded systems [77].

## 4 FAULT MITIGATION METHODS

This section explores "most common" fault mitigation methods. Subsection A examines repetition execution, subsection B discusses Lockstep, and subsection C goes into an in-depth examination of CFC methods, including AUTOSAR, CFC for Permanent faults, CFC for transient faults, and Control-flow Integrity Techniques for Soft Errors-security. Additionally, data integrity is addressed in this section.

The organization of this section is visually represented in Figure 6 providing an overview of each section's focus.

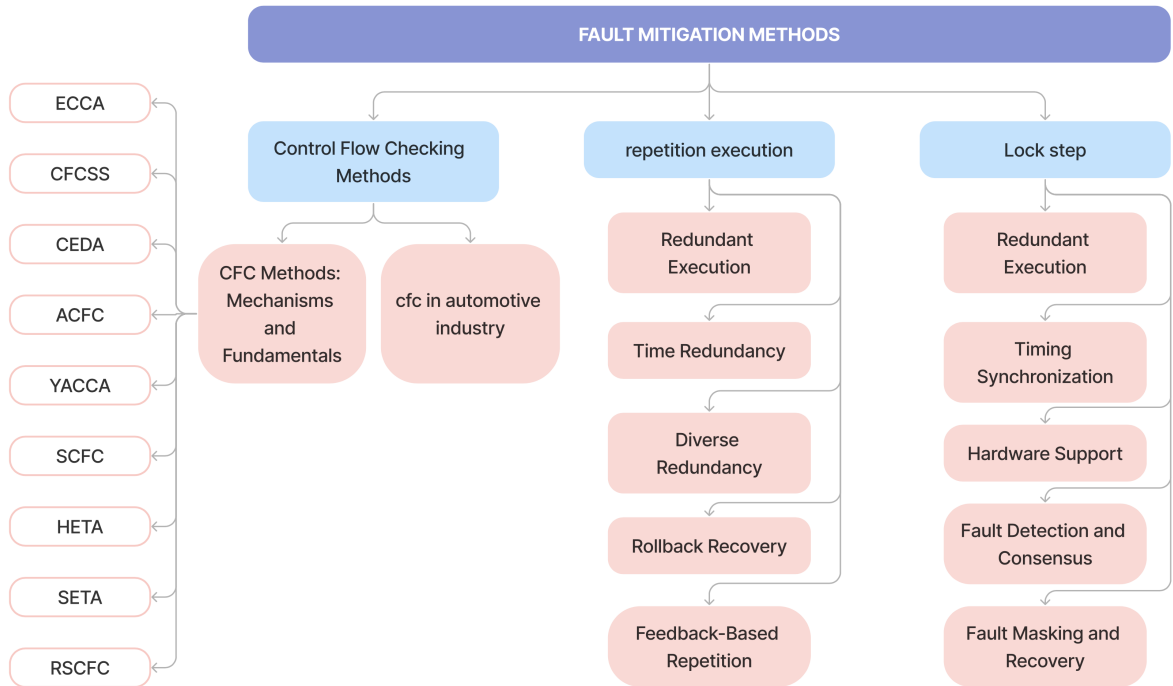

Fig. 6. Fault mitigation methods.

## 4.1 In-Depth Examination of Control Flow Checking Methods

Various techniques have been proposed in the literature to address transient and permanent faults in different parts of a system, targeting both hardware and software components and relying on different forms of redundancy. Among these techniques, CFC stands out as it can cover faults affecting memory components containing the executable program, as well as the hardware components handling the program and its flow [93]. CFC has been suggested to handle reliability issues for both transient and permanent faults [21, 98], and more recently, it has been applied to address security issues caused by the injection of malicious faults [6, 32]. Malicious faults, within the context of fault tolerance, refer to deliberate and intentional actions taken by malicious actors to disrupt or compromise the normal functioning of a computer system, network, or software application. These actions are aimed at exploiting vulnerabilities in order to compromise the system's integrity, availability, or confidentiality [16]. Unlike transient and permanent faults, which often arise from natural hardware failures or environmental factors, malicious faults are caused by human intent and typically involve actions such as hacking, malware deployment, or unauthorized access.

In a cost-effective method proposed in [115], transient faults are detected through coarse-grain CFC, achieving efficiency by simplifying signature calculations within BBs and conducting checks at a coarse-grain level. To assess the effectiveness of this approach, a comprehensive fault injection campaign was conducted, using single bit-flips to model transient faults. Transient faults may not cause permanent damage to the hardware, but they can silently corrupt an application's correctness during runtime or even lead to system crashes. For instance, HP [71] reported frequent failures in their 2048-CPU system at the Los Alamos National Laboratory due to high-energy cosmic rays. A study [30] revealed that the BlueGene/L machine installed in Lawrence Livermore National Labs experienced soft errors approximately every four hours. Considering the estimated reliability drop per bit with each generation of processors [25], it becomes essential to provide transient fault protection schemes for both current and future systems. Transient fault detection techniques rely on different forms of redundant checking, either in hardware or software. Hardware solutions like DMR, TMR, and watchdog processors [69] are employed in systems like IBM Z-Series servers [19], HP NonStop system [23], and Boeing 777 airplanes [114]. However, hardware-based solutions introduce unavoidable area and energy costs, making them unsuitable for commodity-embedded systems. Software-based redundant checking, on the other hand, is more appealing for transient fault detection due to its lower production costs and higher flexibility. Securing control flows is crucial for transient fault protection, as CFEs are more likely to cause programs to behave incorrectly. While traditional software methods [76, 106] provide high fault coverage, they inject a significant number of validating instructions into programs, resulting in moderate to large performance overhead. Recent studies [56, 117] attempt to reduce this validation overhead by injecting fewer instructions, but they may sacrifice fault coverage due to their heuristic approaches. Software-based transient fault detection techniques are categorized into data flow protection and control flow protection. Although data flow errors can be masked during program executions, CFEs are more challenging to hide. This work focuses on detecting illegal control flows since they can lead to incorrect program behavior. Researchers from industry and academia have been actively seeking solutions to counter the threat of transient faults in both hardware and software. Hardware-only solutions, with sufficient resources, are more efficient for a single, fixed reliability policy, while software-only solutions offer flexibility and lower costs. Software-only solutions can be deployed immediately on existing hardware by recompiling the application. However, devising correct software solutions for transient faults is a challenging task due to the numerous fault scenarios. Various techniques are suggested in the literature for detecting transient faults, falling into two general classes: hardware or software redundancy. Hardware-based methods provide better fault coverage but impose higher costs and overhead on the system, making them less suitable for some general-purpose applications. On the other hand, software-based techniques offer less fault coverage and larger delay but are more cost-effective, flexible, and applicable to different types of COTS systems.

*4.1.1 CFC Methods: Mechanisms and Fundamentals.* One common approach in CFC methods is signature monitoring, where redundant instructions are inserted into the software unit's source code. This method proves advantageous as it doesn't necessitate any special hardware or operating system requirements, making it adaptable to Commercial off the Shelf (COTS) micro-controllers, even low-power units. Furthermore, CFC harmoniously complements hardware-based hardening techniques, such as watchdogs, and can be expedited through external hardware support for run-time signature execution and comparison.

In the context of CFC, the typical approach involves dividing the source code into BBs and meticulously inspecting the code within these blocks, along with the branches connecting them. To facilitate this process, a watchdog processor can be employed, enabling efficient and effective control flow verification.

The errors analyzed in these methods fall into three general categories: a) illegal jumps within a BB, b) illegal jumps among BBs, and c) illegal jumps from a BB to the unused memory space. These illegal jumps result in Control Flow Errors (CFEs). The following shows six situations that jumps result in a CFE:

- an illegal jump from the end of one BB to the beginning of another BB.
- a legal but incorrect jump from the end of one BB to the beginning of another BB.
- a jump from the end of one BB to any point within another BB.
- a jump from any point within one BB to any point within another BB.
- a jump from any point within a BB to another point within the same block.

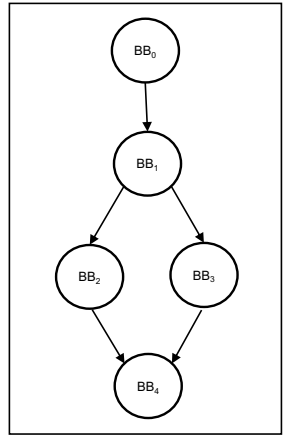

$G = (V, E)$
$V = \{BB_0, BB_1, BB_2, BB_3, BB_4\}$
$E = \{e_0, e_1, e_2, e_3, e_4\}$
$e_0 = \{BB_0, BB_1\}$
$e_1 = \{BB_1, BB_2\}$
$e_2 = \{BB_1, BB_3\}$
$e_3 = \{BB_2, BB_4\}$
$e_4 = \{BB_3, BB_4\}$

| $BB_0$ | `while (int i>10) {` |
|---|---|
| $BB_1$ | `If y==1` |
| $BB_2$ | `    y ++;` |
| $BB_3$ | `Else`<br>`    y=y*2;` |
| $BB_4$ | `i++;`<br>`}` |

Fig. 7. Example of program CFG and sample code. The execution path from BB1 to BB2 or from BB1 to BB3 is valid, while a jump from BB1 to BB4 is invalid and referred to as CFE. [99]

- an illegal jump from a BB to the unused space of memory, which refers to the space between BBs.

It is important to note that regardless of the approach used (software or hardware-based), in industrial applications, the method should be capable of handling the aforementioned errors while minimizing memory overhead and execution time increase.

CFC methods utilize Control Flow Graph (CFG) alongside signatures computed by redundant instructions to detect illegal jumps. The basic idea behind signature-monitoring techniques is to assign a static signature to each BB, along with a dynamic global signature. In all CFC detection methods, each BB is associated with a unique static signature. CFC methods employ precise detection approaches by generating the CFG from high-level language source code, defining the BB signatures and their computation methods. During execution of the hardened software component, the signature values computed at run-time are compared with the predetermined signatures. In case of a mismatch, an error signal is activated to trigger the detection. The CFE detection methods can be divided into hardware-based methods [80],[17], mixed software-hardware methods [9],[116], and software-based methods. The hardware-based methods require additional hardware components to detect CFEs. Figure 7 provides a graphical representation of the CFG for a sample source code developed in the C language.

Some of the most commonly used CFC methods are based on comparing the run-time signature computed value with the expected values assigned to each block at the design or compile time. We will clarify them below to shed more light on the techniques.

In [10] authors proposed the Enhanced Control Flow Checking Using Assertions (ECCA) method. It is an enhanced version of Control-Flow Checking Using Assertions (CCA) [70] that is targeted for real-time distributed systems. ECCA overcomes the limitations of CCA by introducing a new assertion methodthat allows for the detection of control-flow errors that were previously undetectable by CCA. In ECCA, each BB in a program should be given a special numerical identification number. Specific assertions that use the identifiers of the involved BBs check the control flow when the processor executes a new BB. Extending the CCA technique, ECCA methods are able to identify all CFEs between various BBs. Still, ECCA methods are unable to identify errors within a single BB or faults that result in incorrect decisions being made on a conditional branch.

In Control Flow Checking by Software Signatures (CFCSS), which is covered in [76], all branches' destinations are evaluated before they jump, not their sources. A global variable named G is initialized with a program's first BB's signature while it is being executed. In order to determine the difference between the signatures of the source and target blocks, CFCSS uses the XOR function to calculate the target block signature from the source block's signature. By comparing the computed signature with the anticipated one, control flow will be examined. The technique outlined in [76] manually inserts control flow checking assertions. This will be accomplished by beginning each BB with a few instructions. First, setting the outgoing signature variable after checking the incoming signature variable. This makes it possible to confirm the accuracy of the execution flow. It does not require specialized hardware, such as a CFC watchdog. It implies that CFCSS is applicable even in the absence of multitasking support by the operating system. If several BBs merge into a single BB at their ends, CFCSS cannot identify errors.

The authors of [106] proposed Control-flow error detection using assertions (CEDA) by assigning a signature verification at the start and end of each BB, and detecting the "aliasing errors" by maintaining unique signatures for each of the aliased blocks. Run-time signatures, which are inserted during compilation, are used by CEDA to identify errors in the control flow effectively. As a result, CEDA can identify all errors that violate the program flow graph, but it cannot identify illegal but correct jumps (according to the program flow graph). As a result, CEDA is unable to detect all the faults.

According to [107], Assertions for Control Flow Checking (ACFC) is a classification scheme for control flow faults and a CFC method that does not rely on predecessor-successor relationships between BBs. The method uses fewer instructions than earlier techniques. Consequently, the method has less memory overhead than the earlier techniques, but its detection performance suffers as a lot.

A CFC technique described in [44] is "Yet Another Control-Flow Checking using Assertions" (YACCA). In this technique, each BB entry and exit point receives a special signature. The benefit of this approach is that it allows for the detection of CFEs that occurred when the program flow changed from one BB's inside to that of one of its legitimate successors, even if the succeeding BB returns control to the BB that was subjected to the incorrect jump. This is possible because the signature is re-evaluated prior to each branch instruction to eliminate the CFE for the incorrect successor. In comparison to CFCSS, the YACCA has higher performance overhead and fewer undetected errors.

Reference [12] proposed Software-Based Control Flow Checking (SCFC). The method makes use of two run-time variables: one that holds the run-time values ID of the BBs and another that holds the run-time signature S. The compile-time signature is created using the same method as SEDSR [13]. In the BB, a CFE can be found in either the run-time ID or the run-time signature S that has the incorrect value. The compile-time value of the BB should be included in the ID, and the predecessor BB's signature should be included in the S. In the BB, ID and S are updated at various locations. After confirming it, the S is updated in the middle of the BB, and the ID is updated to the compile-time id of the succeeding block.

Another approach is Hybrid Error-detection Technique using Assertions (HETA) [17]. HETA can detect incorrect jumps during the program execution. HETA develops CEDA techniques and associates them with hardware resources, a watchdog, for achieving complete fault detection. Using HETA methods cannot detect 100% of the errors.

An alternative approach to detect CFEs in processors without hardware-implemented hardening techniques is the Software-only Error-detection Technique using Assertions (SETA) [34]. This method aims to reduce computation units' costs by utilizing two previously described techniques: Hardware-Enabled Timer-based Assertion (HETA) and Control-flow Error Detection Analysis (CEDA). Both techniques utilize run-time signatures to identify errors related to the control flow. Signatures are calculated in advance and compared with the signatures computed at run-time. To implement SETA, the application code is divided into BBs, and two types of BBs are defined: Type A and Type X. Type A BBs have multiple predecessors, at least one of which has multiple successors. BBs that do not meet these conditions are classified as Type X. The defined BBs are then grouped into networks, where BBs sharing a common predecessor belong to the same network. Each BB has two signatures: the Node Ingress Signature (NIS) and the Node Exit Signature (NES). The NIS is compared when entering the BB, while the NES is checked when exiting the BB. The NIS describes the current BB, while the NES is used to identify the successor network and its subsequent legal successor BBs.

Another technique proposed is the Relationship Signatures for Control Flow Checking (RSCFC) [65]. RSCFC encodes control flow relations between different BBs into specially formatted signatures and inserts CFC instructions at the head and end of every BB. This technique detects inter-block CFEs using three variables: a compile-time signature (si), the CFG locator (Li), and the cumulative signature (mi). RSCFC has a higher fault detection rate compared to CFCSS, but it incurs a higher performance overhead.

In summary, signature monitoring methods such as YACCA [44], CFCSS [76], CEDA [106], RASM [102], SEDSR [13], and ECCA [10] focus on monitoring run-time signatures with compile-time signatures at the BB level to address illegal inter-block jumps during application execution. These methods differ in how signatures are computed and checks are performed. To enhance the existing methods that cover illegal intra-block jumps, instruction monitoring techniques have been developed. These include RSCFC [65], Software implemented error detection (SIED) [74], and Random Additive Control Flow Error Detection (RACFED) [104], which inspects the correct execution order of instructions. Additionally, a behavior-based software technique [67] and the Software Implemented Hardware Fault Tolerance (SIHFT) [43] approach have been presented for detecting CFEs in multi-core architectures and low-cost embedded systems used in safety-critical applications. SIHFT is especially suitable for applications where availability and execution speed are not major concerns.

Table 2 compares the detection coverage and overheads of different CFC methods. The measurements were made by [104] and [103] on implementations at the assembly level. The authors used their software-implemented fault injection (SWIFI) tool to validate the comparisons between the techniques.

Table 2. Compare Control Flow Control techniques [99].

| CFC Method | Used Variables | Signatures | Intra-block | Detection Performance [%] | Code Size Overhead [%] | Execution Time Overhead [%] |
|---|---|---|---|---|---|---|
| **ECCA** | 4 | prime-numbers | $\chi$ | 73.5 | 36.0 | 244.8 |
| **CFCSS** | 2 | randomized-bit | $\chi$ | 75.8 | 15.2 | 76.6 |
| **YACCA** | 2 | bit-field | $\chi$ | 82.8 | 30.0 | 203.2 |
| **RSCFC** | 2 | bit-field | ✓ | 49.4 | 17.5 | 86.8 |
| **SEDSR** | 3 | bit-field | ✓ | 46.8 | 12.3 | 67.1 |
| **SCFC** | 3 | bit-field | ✓ | 60.4 | 22.9 | 115.7 |
| **SIED** | 2 | random numbers | ✓ | 52.4 | 14 | 115.7 |
| **RACFED** | 3 | random numbers | ✓ | N.A. | N.A. | 81.5 |

*4.1.2 cfc in automotive industry.* A good example of applying CFC methods in software architecture, is in the automotive industry. To establish a standardized software architecture for automotive products among manufacturers and suppliers, a framework as a set of specifications has been developed. This framework is named as the AUTomotive Open System ARchitecture (AUTOSAR). AUTOSAR adopts a modular software architecture with standardized interfaces and a runtime environment, effectively segregating application-level software components from underlying software modules and hardware. Such standardization ensures interoperability, reusability, portability, and scalability in AUTOSAR-compliant products, which holds great significance in the industry. Moreover, AUTOSAR places a high priority on functional safety, aligning well with the upcoming ISO 26262 standard. Within the AUTOSAR framework, a methodcalled Watchdog Manager (WdM) incorporates logical monitoring to detect CFEs in program instructions. However, it is worth noting that the specification does not cover the concept of a Control Flow Graph, which limits its potential for comprehensive implementation. Despite this limitation, the AUTOSAR platform has emerged as a prevalent open industry standard for developing in-vehicular systems, particularly in response to the increasing complexity of modern vehicular systems. By offering a modular software architecture that adheres to standardized interfaces and a runtime environment, AUTOSAR effectively separates application-level software components from the underlying basic software modules and physical hardware. To maintain a well-structured framework, AUTOSAR system specifications are stored in the standardized AUTOSAR XML (ARXML) format. However, while the AUTOSAR development process facilitates the monitoring of control flow and timing properties at a low abstraction level, it lacks support for modeling complex monitoring functionality at the software component level. This limitation warrants further consideration when implementing CFC in the context of AUTOSAR's architecture [2, 3].

## 4.2 Repetition Execution

Repetition execution is a widely employed fault mitigation methodin fault-tolerant embedded systems. By executing critical tasks multiple times and comparing the results, repetition execution aims to detect and tolerate faults that may occur during program execution. This subsection provides an overview of repetition execution techniques and their effectiveness in mitigating faults.

- Redundant Execution: One approach to repetition execution involves redundant execution, where critical tasks are executed multiple times in parallel. The results obtained from each execution are compared, and a majority voting or a consensus-based decision methodis used to determine the correct result [46]. Redundant execution techniques can mitigate both permanent and transient faults, ensuring the system's resilience to unexpected failures.
- Time Redundancy: In time-redundant execution, critical tasks are executed at different time instances, providing redundancy in the temporal domain. By repeating the execution of tasks at periodic intervals, fault detection and recovery mechanisms can be incorporated [24]. Time redundancy is particularly effective in mitigating transient faults that may occur intermittently.
- Rollback Recovery: Rollback recovery is a technique that combines repetition execution with checkpointing. Periodically, checkpoints are taken to capture the system's state. In the event of a fault, the system can roll back to a previous checkpoint and re-execute the tasks from that point to ensure correctness and consistency [84]. Rollback recovery provides fault tolerance and can handle permanent faults that affect the system's state.
- Diverse Redundancy: Diverse redundancy is a technique that combines repetition execution with diversity in the implementation or design of critical tasks. Multiple versions of the same task are executed in parallel, each using a different algorithm, implementation, or platform.

By incorporating diversity, the system can tolerate faults that affect only a subset of the redundant tasks, ensuring higher reliability and fault tolerance.
- Feedback-Based Repetition: Feedback-based repetition execution involves continuously monitoring the system's behavior and adapting the repetition methodaccordingly. Fault detection mechanisms analyze the system's output and dynamically adjust the repetition execution parameters, such as the number of repetitions or the timing of task execution, to optimize fault tolerance [8]. This approach improves the system's resilience by adapting to changing fault conditions.

An example of leveraging repetition execution fault mitigation methods is in the automotive domain, where ISO26262 acknowledges recovery through repetition as an accepted error-handling method. This approach involves resetting the specific hardware components involved in a faulty execution and re-executing the affected software components, as described in the AUTOSAR standard for automotive software design. Furthermore, ISO26262-6:2011 clause 10.4.3 states that when generating test cases for software resource usage testing, it is essential to determine the maximum execution time of the program under analysis to demonstrate the schedulability of the integrated system [1].

Taking a repetition execution approach can help enhance the CFC methods, in accurately and efficiently solving detection problems within BBs and across procedures, as well as addressing the issues of control flow error detection hysteresis and reducing time overhead. Existing control flow error detection methods based on signature analysis have limitations due to their reliance on a single type of signature, struggling to balance program residual failure rate and time overhead. To overcome these limitations, a proposed technique called basic block repetition is introduced [75], which involves executing a program multiple times while monitoring the behavior of its BBs to identify anomalies or deviations in control flow, signaling the presence of errors. The process includes instrumentation, execution, monitoring, and analysis of BBs. BB repetition can be detected using various techniques, such as static analysis with hash tables or control flow graphs and dynamic analysis using tracing tools to track executed BBs. This approach offers valuable insights into the control flow dynamics of software systems, aiding in the identification and debugging of CFEs ultimately improving software reliability, security, and performance.

In summary, repetition execution is a prominent fault mitigation methodin fault-tolerant embedded systems. By executing critical tasks multiple times and comparing the results, repetition execution techniques can effectively detect and tolerate faults.

### 4.3 Lock step

Lockstep is a widely used fault mitigation methodin fault-tolerant embedded systems. It involves redundant execution of critical tasks in parallel. The redundant executions are kept synchronized to ensure consistency and fault tolerance. This subsection provides an overview of the lockstep approach for fault mitigation [5, 45, 83, 110].

- Redundant Execution: In the lockstep approach, critical tasks are redundantly executed in parallel with multiple identical copies or processors. These copies execute the same instructions simultaneously, and the outputs are compared to detect discrepancies caused by faults [47]. Lockstep execution is particularly effective in mitigating permanent faults that affect the consistent behavior of a system.
- Fault Detection and Consensus: Lockstep execution relies on fault detection mechanisms to identify inconsistencies among redundant copies. By comparing the outputs of redundant tasks, fault detection algorithms can detect faults and initiate recovery actions. Consensus-based techniques, such as majority voting or Byzantine fault-tolerance algorithms, are commonly used to determine the correct output when discrepancies arise [108].
- Fault Masking and Recovery: The redundant nature of lockstep execution provides fault masking capabilities. If a fault occurs in one copy, the correct output can still be obtained by comparing the results of the other copies. This fault-masking property enhances the fault tolerance of embedded systems. In case of a fault, recovery mechanisms can be triggered to restore the system to a consistent state [18].
- Timing Synchronization: Synchronization of the redundant copies is crucial in lockstep execution to maintain consistency. Precise timing synchronization is required to ensure that the copies execute instructions at the same rate and in the same order. Time-triggered protocols, clock synchronization techniques, or global time references are employed to achieve timing synchronization among the redundant copies [91].
- Hardware Support: Hardware-level support plays a vital role in implementing the lockstep approach efficiently. Specialized hardware architectures and components, such as redundant processors, comparators, and error detection circuits, are designed to facilitate lockstep execution. These hardware features ensure synchronized execution, fault detection, and fault recovery in a timely and efficient manner [51, 82].

All in all, lockstep execution is a powerful fault mitigation method in fault-tolerant embedded systems. By redundantly executing critical tasks in parallel and comparing the results, lockstep ensures fault detection, fault tolerance, and system recovery.

## 5 A NOTE ON CONTROL-FLOW INTEGRITY TECHNIQUES FOR SOFT ERRORS - SECURITY

Control-flow integrity (CFI) techniques are employed to ensure that a program functions as intended without being affected by soft errors, which can arise from external factors like radiation, power surges, or electromagnetic disturbances. These errors have the potential to cause unintended consequences, such as data loss, diminished system reliability, and even security breaches [35]. The primary purpose of CFI techniques is to mitigate the risk of security breaches resulting from soft error-induced deviations by implementing a set of rules on the program's control-flow graph (CFG). This graph represents the program's control flow and the relationships between its various components. These rules dictate the permissible execution paths and prevent any unauthorized or malicious alterations to the control flow. One commonly used CFI technique is "strict control flow integrity" (SCFI), which enforces rules to maintain the integrity of the program's control flow graph during execution. Any attempt to deviate from this graph is detected and prevented, thus safeguarding the program's integrity. Additional CFI techniques include shadow-stack-based CFI, implicit CFI, and hybrid CFI. Soft errors can affect the direct as well as indirect branches, and hence CFI, as is, is not directly applicable for soft errors. Though direct branches can also be protected in a manner similar to dynamic branches, but the already high overhead (20%-60% for dynamic branches only) would become prohibitive [4].

In summary, CFI techniques serve as a set of measures to protect software systems from security breaches caused by soft errors. By enforcing strict rules on the program's control-flow graph, these techniques can identify and prevent any unauthorized or malicious changes to the program's execution, thereby bolstering its security and reliability.

## 5.1 Data Integrity

Data integrity refers to the concept of ensuring that data remains accurate, consistent, and reliable throughout its entire life cycle. In the context of soft error security, data integrity becomes especially crucial in protecting against potential vulnerabilities and risks posed by transient faults or soft errors. These errors can be caused by various factors, such as cosmic radiation, electrical noise, or electromagnetic interference, and can adversely impact the integrity of stored data. To mitigate such risks, data integrity measures involve implementing error detection and correction techniques, such as checksums and parity bits, to detect and correct any errors that may occur.

Maintaining data integrity is relatively straightforward in a standalone system with a single database. This is achieved through the use of database constraints and transactions, typically managed by a database management system (DBMS). Transactions should adhere to the ACID principles (atomicity, consistency, isolation, and durability) to ensure data integrity. Most databases support ACID transactions, which aids in preserving data integrity. However, data integrity in cloud-based systems refers to the preservation of data accuracy. It is crucial to ensure that data remains unchanged and is not lost due to unauthorized user actions. Data integrity forms the foundation for cloud computing services like Software as a service (SaaS), Platform as a Service (PaaS), and Infrastructure as a Service (IaaS) [31]. In addition to storing large volumes of data, cloud environments typically offer data processing services. methods such as RAID-like methods and digital signatures can be employed to maintain data integrity in cloud systems.

Remote verification of data integrity in the cloud is a prerequisite for deploying applications. Bowers et al. introduced the "Proofs of Retrievability" theoretical framework, which combines error correction codes and spot-checking to facilitate remote data integrity checks [29]. The High-Availability and Integrity Layer (HAIL) system utilizes the Proofs of Retrievability (POR) method to verify data storage across different clouds, ensuring redundancy of copies and enabling availability and integrity checks [28]. Schiffman et al. proposed the use of Trusted Platform Modules (TPM) for remote data integrity checks [96].

Due to numerous entities and access points in a cloud environment, authorization plays a vital role in ensuring that only authorized entities interact with data. By preventing unauthorized access, organizations can have greater confidence in data integrity. Monitoring mechanisms provide increased visibility, enabling the identification of any alterations made to data or system information that may affect its integrity. While cloud computing providers are entrusted with maintaining data integrity and accuracy, it is important to establish a third-party supervision method alongside users and cloud service providers.

In summary, data integrity is paramount for safeguarding data accuracy and consistency, particularly in the context of transient faults or soft errors. Techniques like checksums and parity bits are used to detect and correct errors. While standalone systems can ensure data integrity through database constraints and ACID transactions, cloud-based systems require remote verification methods, such as Proofs of Retrievability and Trusted Platform Modules, to maintain data accuracy across various access points. Authorization and monitoring mechanisms also play crucial roles in preserving data integrity in the cloud, requiring collaboration among users, providers, and third-party oversight for effective security.

## 6 EVALUATE THE DEPENDABILITY

The evaluation of dependability in fault-tolerant systems is explored in this section. Figure 8 illustrates the organization of this section. It encompasses fault injection techniques, fault simulation techniques, and fault diagnosis techniques. It has provided essential insights into the dependability of numerous systems and has sparked extensive research in various areas. Several mature fault injection tools have been developed, some of which have been successfully implemented in industrial settings [22, 111]. However, there is a prevailing perception in other communities that fault injection is a solved research problem, with the remaining challenges being primarily engineering-related. Nevertheless, fault injection has been a subject of research for many years [86, 118].

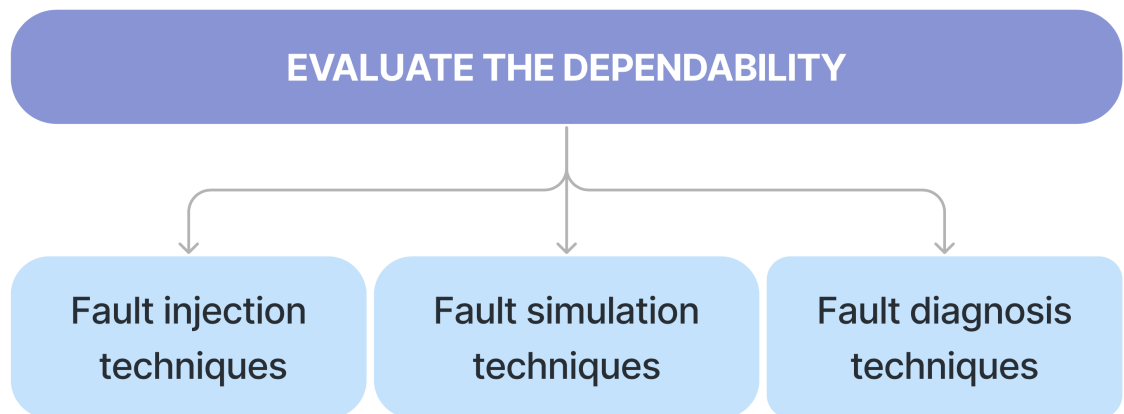

Fig. 8. Evaluate the dependability.

Over time, numerous techniques have been proposed to assess different types of fault effects and analyze faulty behavior. Nonetheless, as technology has advanced, the demands for fault injection have become increasingly rigorous. More complex devices necessitate higher performance to conduct larger fault injection campaigns within an acceptable timeframe [118].

Fault injection, the deliberate introduction of faults into a Circuit Under Test (CUT), has proven to be an effective method for evaluating susceptibility to soft errors. This approach enables researchers to introduce faults, thus reducing experiment duration artificially. The goals of fault injection include verifying circuit fault tolerance, predicting circuit behavior in the presence of faults, identifying critical components that require mitigation, and validating mitigation approaches.

Various fault injection methods have been proposed, with physical fault injection utilizing particle accelerators being the most widely accepted. However, this method is costly and suitable only for certified circuits. Other physical techniques involve the use of lasers or electromagnetic interference. Alternatively, logical fault injection can be employed by altering register or memory contents within the CUT and observing the effects. This method is simpler and more cost-effective to implement, but it has limitations as it may not provide access to all circuit components. Additional techniques are needed for effective guidance and validation of mitigation techniques during the design phase. These techniques involve evaluating mitigation needs and effectiveness using logical fault injection on a design model. For this purpose, simulation, emulation, or a combination of both, along with appropriate fault models and design tools, can be utilized. The bit-flip model is commonly utilized for logical emulation of radiation-induced faults, allowing for the injection of single and multiple faults. Table 3 provides a summary of how the bit-flip model can be used to model single and multiple effects.

Fault injection necessitates a suitable CUT model, and the level of detail depends on the type of fault being considered. Performance is a critical factor in fault injection, as a sufficient number of faults must be injected to achieve statistical significance. Recent developments have focused on enhancing the performance of fault injection to enable larger campaigns and support more comprehensive fault analysis. [49]

- Fault injection techniques
  Fault injection is a testing and debugging technique that deliberately introduces faults into memory and register circuits to observe and evaluate their behavior and response under faulty conditions. The level of abstraction at which fault injection is performed, such as hardware, software, or firmware, depends on the type and complexity of the circuit. The stage of fault injection, whether pre-silicon, post-silicon, or in-field, is determined by the availability and accessibility of the circuit. The techniques for fault injection vary based on the fault type, model, level, and stage. Common fault injection techniques for memory and register circuits include physical, electrical, optical, thermal, magnetic, radiation, software, or firmware fault injection. [22]
- *Fault simulation techniques:* Fault simulation is a testing and debugging technique that utilizes software models or emulators to simulate the behavior and effects of faults in memory and register circuits without physically altering the hardware components. The level at which fault simulation is conducted, such as functional, structural, or behavioral, depends on the type and complexity of the circuit. The stage of fault simulation, whether pre-silicon, post-silicon, or in-field, is determined by the availability and accessibility of the circuit. The techniques for fault simulation vary based on the fault type, model, level, and stage. Common fault simulation techniques for memory and register circuits include fault list, fault dictionary, fault coverage, fault equivalence, fault collapsing, or fault grading [87].
- *Fault diagnosis techniques:* Fault diagnosis is a debugging technique that utilizes testing and analysis methods to identify and locate faulty components or regions in memory and register circuits. The level at which fault diagnosis is performed, such as gate, transistor, or layout, depends on the type and complexity of the circuit. The stage of fault diagnosis, whether pre-silicon, post-silicon, or in-field, is determined by the availability and accessibility of the circuit. The techniques for fault diagnosis vary based on the fault type, model, level, and stage. Common fault diagnosis techniques for memory and register circuits include signature analysis, parity check, checksum, syndrome decoding, error correction code (ECC), error detection and correction (EDAC), or fault localization [36].

Table 3. Models of soft errors for fault injection. Single-Event Upset (SEU), Multiple-Cell Upset (MCU), Single-Event Multiple Transient (SEMT), Single-Event Transient (SET) [38].

| Feature | SEU/MCU | SET/SEMT |
|---|---|---|
| Effect | Single/multiple bit-flip | Single/multiple bit-flip |
| Where? | Any flip-flop | Any gate |
| When? | Any clock cycle | Any time |
| For how long? | 1 clock cycle (typically) | Variable pulse width |

## 7 CONCLUSION

In this paper, we explored and surveyed various fault-tolerance methods designed to mitigate random hardware failures in embedded systems, focusing on real-time embedded systems. The increasing use of embedded systems in safety or mission-critical applications necessitates advanced and reliable fault tolerance methods to ensure seamless automation and operational efficiency in commercial and industrial contexts. The significance of fault tolerance in modern computing systems was discussed, and insights were given on how fault tolerance techniques enhance system dependability by masking faults and detecting errors, allowing for uninterrupted service provision in the presence of internal faults. The paper has highlighted the differences in using hardware or software redundancy to achieve fault tolerance goals, ensuring reliable system functioning.

Table 4. Summary of Sections

| Section/Subsection | Content | Key Points |
|---|---|---|
| **Background** | <ul><li>Register File</li><li>Integer Unit (IU) and Floating Point Unit (FPU)</li><li>Bus Unit</li><li>Control Unit</li><li>Debug Unit</li><li>Instruction Cache</li><li>Data Cache</li></ul> | <ul><li>Impact of SEUs on different components of a system.</li></ul> |
| **Designing Fault-Tolerant Systems** | <ul><li>Hardware-Based</li><li>Software-Based</li><li>Hybrid-Based</li></ul> | <ul><li>Description of hardware and software fault-tolerance techniques.</li></ul> |
| **Fault Mitigation Methods** | <ul><li>Fault Mitigation Methods</li><li>Control Flow Checking Methods</li><li>CFC Methods: Mechanisms and Fundamentals</li><li>CFC in Automotive Industry</li><li>Repetition Execution</li><li>Lockstep</li></ul> | <ul><li>Overview of common fault mitigation methods.</li><li>Detailed exploration of control flow checking methods and their application in the automotive industry.</li><li>Overview of repetition execution and lockstep as fault mitigation methods.</li></ul> |
| **Control-flow Integrity Techniques** | <ul><li>Control-flow Integrity Techniques for Soft Errors - security</li><li>Data Integrity</li></ul> | <ul><li>Overview of control-flow integrity techniques for preventing security breaches caused by soft errors.</li><li>Importance of data integrity in protecting against vulnerabilities posed by soft errors, especially in cloud environments.</li></ul> |
| **Evaluate the Dependability** | <ul><li>Evaluate the Dependability</li><li>Fault injection techniques</li><li>Fault simulation techniques</li><li>Fault diagnosis techniques</li><li>Models of soft errors for fault injection</li></ul> | <ul><li>Exploration of fault injection, simulation, and diagnosis techniques for evaluating dependability in fault-tolerant systems.</li><li>Summary of models for soft errors in fault injection.</li></ul> |

Special attention has been given to software fault tolerance, as software faults are a leading cause of system failures. While software engineering endeavors to remove most deterministic design faults, it is practically impossible to guarantee that complex software designs are entirely free of such faults. Hence, software fault tolerance techniques are employed as an additional layer of protection to ensure continued service at an acceptable level of performance and safety. Moreover, the increasing complexity and optimization of computer systems for price and performance have introduced the challenge of soft errors or transient bit-errors. This challenge emphasizes the critical role of fault tolerance in modern computing systems, as these errors can potentially lead to system malfunctioning.

The survey has covered various fault tolerance techniques, including hardware, software, and hybrid redundancy, providing valuable insights into their benefits and applicability in different contexts. Additionally, we have discussed fault-tolerance approaches tailored specifically for resource-constrained embedded systems, acknowledging the importance of considering limited memory and low-end computation environments in such systems. Nevertheless, more research and development in fault-tolerance methods is needed, particularly in the realm of real-time embedded systems, to ensure the reliable and resilient operation of interconnected computing systems.

Overall, this survey paper provides valuable insights into fault mitigation techniques and emphasizes the significance of fault tolerance in ensuring the dependability and functionality of modern computing systems. The presented methods, such as CFC, redundancy approaches, optimized resource management, and security-oriented measures, pave the way for further advancements in the field of fault tolerance and its application in critical computing systems.

The key insights from each section are succinctly summarized in Table 4, providing a quick reference for readers to grasp the essential content discussed throughout this survey.

## REFERENCES

[1] 2018. ISO 26262:2018 Road Vehicles – Functional Safety.
[2] 2022. AUTOSAR: Specification of Operating System (2011),. https://www.autosar.org/fileadmin/standards/R20-11/CP/AUTOSAR_SWS_OS.pdf.
[3] 2022. AUTOSAR: Specification of Watchdog Manager (2011),. https://www.autosar.org/fileadmin/standards/R21-11/CP/AUTOSAR_SWS_WatchdogManager.pdf.
[4] Martín Abadi, Mihai Budiu, Ulfar Erlingsson, and Jay Ligatti. 2009. Control-flow integrity principles, implementations, and applications. *ACM Transactions on Information and System Security (TISSEC)* 13, 1 (2009), 1–40.
[5] F. Abate, L. Sterpone, and M. Violante. 2007. A new mitigation approach for soft errors in embedded processors. In *2007 9th European Conference on Radiation and Its Effects on Components and Systems*. 1–6. https://doi.org/10.1109/RADECS.2007.5205504
[6] Jacob A Abraham and Ramtilak Vemu. 2009. Control flow deviation detection for software security. US Patent App. 12/484,839.
[7] Innocent Agbo, Mottaqiallah Taouil, Said Hamdioui, Pieter Weckx, Stefan Cosemans, Francky Catthoor, and Wim Dehaene. 2016. Read path degradation analysis in SRAM. In *2016 21th IEEE European Test Symposium (ETS)*. 1–2. https://doi.org/10.1109/ETS.2016.7519325
[8] Ali Sharif Ahmadian, Mahdieh Hosseingholi, and Alireza Ejlali. 2010. A control-theoretic energy management for fault-tolerant hard real-time systems. In *2010 IEEE International Conference on Computer Design*. IEEE, 173–178.
[9] Ryan Ahmed, Mohammed El Sayed, S Andrew Gadsden, Jimi Tjong, and Saeid Habibi. 2014. Automotive internal-combustion-engine fault detection and classification using artificial neural network techniques. *IEEE Transactions on vehicular technology* 64, 1 (2014), 21–33.
[10] Zeyad Alkhalifa, VS Sukumaran Nair, Narayanan Krishnamurthy, and Jacob A. Abraham. 1999. Design and evaluation of system-level checks for on-line control flow error detection. *IEEE Transactions on Parallel and Distributed Systems* 10, 6 (1999), 627–641.
[11] Haleh Ardebili, Jiawei Zhang, and Michael G. Pecht. 2019. 10 - Trends and challenges. In *Encapsulation Technologies for Electronic Applications (Second Edition)* (second edition ed.), Haleh Ardebili, Jiawei Zhang, and Michael G. Pecht

(Eds.). William Andrew Publishing, 431–479. https://doi.org/10.1016/B978-0-12-811978-5.00010-9

[12] Seyyed Amir Asghari, Hassan Taheri, Hossein Pedram, and Okyay Kaynak. 2013. Software-based control flow checking against transient faults in industrial environments. *IEEE Transactions on Industrial Informatics* 10, 1 (2013), 481–490.

[13] Seyyed Amir Asghari, Hassan Taheri, Hossein Pedram, and Okyay Kaynak. 2013. Software-based control flow checking against transient faults in industrial environments. *IEEE Transactions on Industrial Informatics* 10, 1 (2013), 481–490.

[14] Todd M Austin. 1999. DIVA: A reliable substrate for deep submicron microarchitecture design. In *MICRO-32. Proceedings of the 32nd Annual ACM/IEEE International Symposium on Microarchitecture*. IEEE, 196–207.

[15] Algirdas Avizienis. 1995. The methodology of n-version programming. *Software fault tolerance* 3 (1995), 23–46.

[16] Algirdas Avizienis, J-C Laprie, Brian Randell, and Carl Landwehr. 2004. Basic concepts and taxonomy of dependable and secure computing. *IEEE transactions on dependable and secure computing* 1, 1 (2004), 11–33.

[17] Jose Rodrigo Azambuja, Mauricio Altieri, Jürgen Becker, and Fernanda Lima Kastensmidt. 2013. HETA: Hybrid error-detection technique using assertions. *IEEE Transactions on Nuclear Science* 60, 4 (2013), 2805–2812.

[18] Massimo Baleani, Alberto Ferrari, Leonardo Mangeruca, Alberto Sangiovanni-Vincentelli, Maurizio Peri, and Saverio Pezzini. 2003. Fault-tolerant platforms for automotive safety-critical applications. In *Proceedings of the 2003 international conference on Compilers, architecture and synthesis for embedded systems*. 170–177.

[19] Wendy Bartlett and Lisa Spainhower. 2004. Commercial fault tolerance: A tale of two systems. *IEEE Transactions on dependable and secure computing* 1, 1 (2004), 87–96.

[20] Robert C Baumann. 2005. Radiation-induced soft errors in advanced semiconductor technologies. *IEEE Transactions on Device and materials reliability* 5, 3 (2005), 305–316.

[21] Alfredo Benso, Stefano Di Carlo, Giorgio Di Natale, and Paolo Prinetto. 2003. A watchdog processor to detect data and control flow errors. In *9th IEEE On-Line Testing Symposium, 2003. IOLTS 2003.* IEEE, 144–148.

[22] Alfredo Benso and Paolo Prinetto. 2003. *Fault injection techniques and tools for embedded systems reliability evaluation.* Vol. 23. Springer Science & Business Media.

[23] David Bernick, Bill Bruckert, Paul Del Vigna, David Garcia, Robert Jardine, Jim Klecka, and Jim Smullen. 2005. NonStop/spl reg/advanced architecture. In *2005 International Conference on Dependable Systems and Networks (DSN'05)*. IEEE, 12–21.

[24] Vladimir A Bogatyrev and AV Bogatyrev. 2015. Functional reliability of a real-time redundant computational process in cluster architecture systems. *Automatic Control and Computer Sciences* 49 (2015), 46–56.

[25] Shekhar Borkar et al. 2004. Microarchitecture and design challenges for gigascale integration. In *MICRO*, Vol. 37. 3–3.

[26] Shekhar Borkar, Norman P. Jouppi, and Per Stenstrom. 2007. Microprocessors in the Era of Terascale Integration. In *2007 Design, Automation & Test in Europe Conference & Exhibition.* 1–6. https://doi.org/10.1109/DATE.2007.364597

[27] N.S. Bowen and D.K. Pradham. 1993. Processor- and memory-based checkpoint and rollback recovery. *Computer* 26, 2 (1993), 22–31. https://doi.org/10.1109/2.191981

[28] Kevin D. Bowers, Ari Juels, and Alina Oprea. 2009. HAIL: A High-Availability and Integrity Layer for Cloud Storage. In *Proceedings of the 16th ACM Conference on Computer and Communications Security* (Chicago, Illinois, USA) *(CCS '09)*. Association for Computing Machinery, New York, NY, USA, 187–198. https://doi.org/10.1145/1653662.1653686

[29] Kevin D Bowers, Ari Juels, and Alina Oprea. 2009. Proofs of retrievability: Theory and implementation. In *Proceedings of the 2009 ACM workshop on Cloud computing security*. 43–54.

[30] Greg Bronevetsky, B de Supinski, and Martin Schulz. 2009. *A foundation for the accurate prediction of the soft error vulnerability of scientific applications.* Technical Report. Lawrence Livermore National Lab.(LLNL), Livermore, CA (United States).

[31] K Chandrasekaran. 2014. *Essentials of cloud computing.* CrC Press.

[32] Ameya Chaudhari, Junyoung Park, and Jacob Abraham. 2013. A framework for low overhead hardware based runtime control flow error detection and recovery. In *2013 IEEE 31st VLSI Test Symposium (VTS)*. IEEE, 1–6.

[33] Eduardo Chielle, Boyang Du, Fernanda L. Kastensmidt, Sergio Cuenca-Asensi, Luca Sterpone, and Matteo Sonza Reorda. 2016. Hybrid soft error mitigation techniques for COTS processor-based systems. In *2016 17th Latin-American Test Symposium (LATS)*. 99–104. https://doi.org/10.1109/LATW.2016.7483347

[34] Eduardo Chielle, Gennaro S Rodrigues, Fernanda L Kastensmidt, Sergio Cuenca-Asensi, Lucas A Tambara, Paolo Rech, and Heather Quinn. 2015. S-SETA: Selective software-only error-detection technique using assertions. *IEEE transactions on Nuclear Science* 62, 6 (2015), 3088–3095.

[35] Gilad Dar, Giorgio Di Natale, and Osnat Keren. 2021. Nonlinear code-based low-overhead fine-grained control flow checking. *IEEE Trans. Comput.* 71, 3 (2021), 658–669.

[36] Steven X Ding. 2008. *Model-based fault diagnosis techniques: design schemes, algorithms, and tools.* Springer Science & Business Media.

[37] Elena Dubrova. 2013. *Fault-tolerant design.* Springer.

[38] Luis Entrena, Mario García-Valderas, Almudena Lindoso, Marta Portela-Garcia, and Enrique San Millán. 2019. *Fault Injection Methodologies*. Springer International Publishing, Cham, 127–144. https://doi.org/10.1007/978-3-030-04660-6_6

[39] Bill Fleming. 2011. Microcontroller Units in Automobiles [Automotive Electronics]. *IEEE Vehicular Technology Magazine* 6, 3 (2011), 4–8. https://doi.org/10.1109/MVT.2011.941888

[40] Rémi Gaillard. 2010. Single event effects: Mechanisms and classification. In *Soft errors in modern electronic systems*. Springer, 27–54.

[41] Zhiwei Gao, Carlo Cecati, and Steven X Ding. 2015. A survey of fault diagnosis and fault-tolerant techniques—Part I: Fault diagnosis with model-based and signal-based approaches. *IEEE transactions on industrial electronics* 62, 6 (2015), 3757–3767.

[42] Daniel Gil-Tomás, Joaquín Gracia-Morán, J.-Carlos Baraza-Calvo, Luis-J. Saiz-Adalid, and Pedro-J. Gil-Vicente. 2012. Studying the effects of intermittent faults on a microcontroller. *Microelectronics Reliability* 52, 11 (2012), 2837–2846. https://doi.org/10.1016/j.microrel.2012.06.004

[43] Olga Goloubeva, Maurizio Rebaudengo, M Sonza Reorda, and Massimo Violante. 2003. Soft-error detection using control flow assertions. In *Proceedings 18th IEEE Symposium on Defect and Fault Tolerance in VLSI Systems*. IEEE, 581–588.

[44] Olga Goloubeva, Maurizio Rebaudengo, M Sonza Reorda, and Massimo Violante. 2005. Improved software-based processor control-flow errors detection technique. In *Annual Reliability and Maintainability Symposium, 2005. Proceedings.* IEEE, 583–589.

[45] Julen Gomez-Cornejo, Aitzol Zuloaga, Uli Kretzschmar, Unai Bidarte, and Jaime Jimenez. 2013. Fast context reloading lockstep approach for SEUs mitigation in a FPGA soft core processor. In *IECON 2013 - 39th Annual Conference of the IEEE Industrial Electronics Society*. 2261–2266. https://doi.org/10.1109/IECON.2013.6699483

[46] Kim P Gostelow. 2011. The design of a fault-tolerant, real-time, multi-core computer system. In *2011 Aerospace Conference*. IEEE, 1–8.

[47] Florian Haas. 2019. Fault-tolerant execution of parallel applications on x86 multi-core processors with hardware transactional memory. (2019).

[48] Said Hamdioui, Dimitris Gizopoulos, Groeseneken Guido, Michael Nicolaidis, Arnaud Grasset, and Philippe Bonnot. 2013. Reliability challenges of real-time systems in forthcoming technology nodes. In *2013 Design, Automation & Test in Europe Conference & Exhibition (DATE)*. 129–134. https://doi.org/10.7873/DATE.2013.040

[49] Mei-Chen Hsueh, T.K. Tsai, and R.K. Iyer. 1997. Fault injection techniques and tools. *Computer* 30, 4 (1997), 75–82. https://doi.org/10.1109/2.585157

[50] Krzysztof Iniewski. 2018. *Radiation effects in semiconductors*. CRC press.

[51] Casey M. Jeffery and Renato J. O. Figueiredo. 2012. A Flexible Approach to Improving System Reliability with Virtual Lockstep. *IEEE Transactions on Dependable and Secure Computing* 9, 1 (2012), 2–15. https://doi.org/10.1109/TDSC.2010.53

[52] Fernanda Kastensmidt and Paolo Rech. 2016. Radiation effects and fault tolerance techniques for FPGAs and GPUs. In *FPGAs and Parallel Architectures for Aerospace Applications: Soft Errors and Fault-Tolerant Design*. Springer, 3–17.

[53] Fernanda Lima Kastensmidt, Luigi Carro, and Ricardo Augusto da Luz Reis. 2006. *Fault-tolerance techniques for SRAM-based FPGAs*. Vol. 1. Springer.

[54] Hans G. Kerkhoff and H. Ebrahimi. 2015. Intermittent Resistive Faults in Digital CMOS Circuits. In *2015 IEEE 18th International Symposium on Design and Diagnostics of Electronic Circuits & Systems*. 211–216. https://doi.org/10.1109/DDECS.2015.12

[55] Seyab Khan, Said Hamdioui, Halil Kukner, Praveen Raghavan, and Francky Catthoor. 2012. BTI impact on logical gates in nano-scale CMOS technology. In *2012 IEEE 15th International Symposium on Design and Diagnostics of Electronic Circuits & Systems (DDECS)*. 348–353. https://doi.org/10.1109/DDECS.2012.6219086

[56] Daya Shanker Khudia and Scott Mahlke. 2013. Low cost control flow protection using abstract control signatures. In *Proceedings of the 14th ACM SIGPLAN/SIGBED conference on Languages, compilers and tools for embedded systems*. 3–12.

[57] J.C. Knight. 2002. Safety critical systems: challenges and directions. In *Proceedings of the 24th International Conference on Software Engineering. ICSE 2002*. 547–550.

[58] John C Knight. 2002. Safety critical systems: challenges and directions. In *Proceedings of the 24th international conference on software engineering*. 547–550.

[59] R Koga, SH Penzin, KB Crawford, and WR Crain. 1997. Single event functional interrupt (SEFI) sensitivity in microcircuits. In *RADECS 97. Fourth European Conference on Radiation and its Effects on Components and Systems (Cat. No. 97TH8294)*. IEEE, 311–318.

[60] Philip Koopman. 2010. *Better embedded system software*. Drumnadrochit Education Pittsburgh.

[61] Israel Koren and C Mani Krishna. 2020. *Fault-tolerant systems*. Morgan Kaufmann.

[62] Israel Koren and C Mani Krishna. 2020. *Fault-tolerant systems*. Morgan Kaufmann.

[63] Jean-Claude Laprie, Jean Arlat, Christian Beounes, and Karama Kanoun. 1995. Definition and analysis of hardware- and-software fault-tolerant architectures. In *Predictably Dependable Computing Systems*. Springer, 103–122.

[64] Peter Alan Lee, Thomas Anderson, Peter Alan Lee, and Thomas Anderson. 1990. *Fault tolerance*. Springer.

[65] Aiguo Li and Bingrong Hong. 2007. Software implemented transient fault detection in space computer. *Aerospace science and technology* 11, 2-3 (2007), 245–252.

[66] CA Lisboa, Marcelo Ienczczak Erigson, and Luigi Carro. 2007. System level approaches for mitigation of long duration transient faults in future technologies. In *12th IEEE European test symposium (ETS'07)*. IEEE, 165–172.

[67] Mohammad Maghsoudloo, Hamid R Zarandi, and Navid Khoshavi. 2012. An efficient adaptive software-implemented technique to detect control-flow errors in multi-core architectures. *Microelectronics Reliability* 52, 11 (2012), 2812–2828.

[68] Aamer Mahmood and Edward J McCluskey. 1988. Concurrent error detection using watchdog processors-a survey. *IEEE Trans. Comput.* 37, 2 (1988), 160–174.

[69] Aamer Mahmood and Edward J McCluskey. 1988. Concurrent error detection using watchdog processors-a survey. *IEEE Trans. Comput.* 37, 2 (1988), 160–174.

[70] Lee D McFearin and VS Sukumaran Nair. 1998. Control Flow Checking Using Assertions. *Dependable Computing and Fault Tolerant Systems* 10 (1998), 183–200.

[71] Shubu Mukherjee. 2011. *Architecture design for soft errors*. Morgan Kaufmann.

[72] Victor P. Nelson. 1990. Fault-tolerant computing: Fundamental concepts. *Computer* 23, 7 (1990), 19–25.

[73] Victor F Nicola. 1994. *Checkpointing and the modeling of program execution time*. University of Twente, Department of Computer Science and Department of . . . .

[74] Bogdan Nicolescu, Yvon Savaria, and Raoul Velazco. 2003. SIED: Software implemented error detection. In *Proceedings 18th IEEE Symposium on Defect and Fault Tolerance in VLSI Systems*. IEEE, 589–596.

[75] Motoki Obata, Jun Shirako, Hiroki Kaminaga, Kazuhisa Ishizaka, and Hironori Kasahara. 2005. Hierarchical parallelism control for multigrain parallel processing. In *Languages and Compilers for Parallel Computing: 15th Workshop, LCPC 2002, College Park, MD, USA, July 25-27, 2002. Revised Papers 15*. Springer, 31–44.

[76] Nahmsuk Oh, Philip P Shirvani, and Edward J McCluskey. 2002. Control-flow checking by software signatures. *IEEE transactions on Reliability* 51, 1 (2002), 111–122.

[77] ADRIA BARROS DE OLIVEIRA. 2017. Applying Dual-Core Lockstep in Embedded Processors to Mitigate Radiation- induced Soft Errors. Available at https://www.lume.ufrgs.br/bitstream/handle/10183/173785/001061371.pdf?sequence=1.

[78] Daniel Oliveira, Sean Blanchard, Nathan Debardeleben, Fernando F. Dos Santos, Gabriel Piscoya Dávila, Philippe Navaux, Carlo Cazzaniga, Christopher Frost, Robert C. Baumann, and Paolo Rech. 2020. Thermal Neutrons: a Possible Threat for Supercomputers and Safety Critical Applications. In *2020 IEEE European Test Symposium (ETS)*. 1–6. https://doi.org/10.1109/ETS48528.2020.9131597

[79] Sangwoo Pae, Jose Maiz, Chetan Prasad, and Bruce Woolery. 2008. Effect of BTI Degradation on Transistor Variability in Advanced Semiconductor Technologies. *IEEE Transactions on Device and Materials Reliability* 8, 3 (2008), 519–525. https://doi.org/10.1109/TDMR.2008.2002351

[80] Luis Parra, Almudena Lindoso, Marta Portela, Luis Entrena, Felipe Restrepo-Calle, Sergio Cuenca-Asensi, and Antonio Martínez-Álvarez. 2014. Efficient mitigation of data and control flow errors in microprocessors. *IEEE Transactions on Nuclear Science* 61, 4 (2014), 1590–1596.

[81] Edward Petersen. 2011. *Single event effects in aerospace*. John Wiley & Sons.

[82] M. Peña-Fernández, A. Serrano-Cases, A. Lindoso, S. Cuenca-Asensi, L. Entrena, Y. Morilla, Pedro Martín-Holgado, and A. Martínez-Álvarez. 2022. Hybrid Lockstep Technique for Soft Error Mitigation. *IEEE Transactions on Nuclear Science* 69, 7 (2022), 1574–1581. https://doi.org/10.1109/TNS.2022.3149867

[83] Hung-Manh Pham, Sébastien Pillement, and Stanisław J. Piestrak. 2013. Low-overhead fault-tolerance technique for a dynamically reconfigurable softcore processor. *IEEE Trans. Comput.* 62, 6 (2013), 1179–1192. https://doi.org/10.1109/TC.2012.55

[84] Paul Pop, Viacheslav Izosimov, Petru Eles, and Zebo Peng. 2009. Design optimization of time-and cost-constrained fault-tolerant embedded systems with checkpointing and replication. *IEEE Transactions on Very Large Scale Integration (VLSI) Systems* 17, 3 (2009), 389–402.

[85] Dhiraj K Pradhan et al. 1996. *Fault-tolerant computer system design*. Vol. 132. Prentice-Hall Englewood Cliffs.

[86] Heather M. Quinn, Dolores A. Black, William H. Robinson, and Stephen P. Buchner. 2013. Fault Simulation and Emulation Tools to Augment Radiation-Hardness Assurance Testing. *IEEE Transactions on Nuclear Science* 60, 3 (2013), 2119–2142. https://doi.org/10.1109/TNS.2013.2259503

[87] Heather M Quinn, Dolores A Black, William H Robinson, and Stephen P Buchner. 2013. Fault simulation and emulation tools to augment radiation-hardness assurance testing. *IEEE Transactions on Nuclear Science* 60, 3 (2013), 2119–2142.

[88] Martin Radetzki, Chaochao Feng, Xueqian Zhao, and Axel Jantsch. 2013. Methods for Fault Tolerance in Networks-on-Chip. *ACM Comput. Surv.* 46, 1, Article 8 (jul 2013), 38 pages. https://doi.org/10.1145/2522968.2522976

[89] Brian Randell and Jie Xu. 1995. The evolution of the recovery block concept. *Software fault tolerance* 3 (1995), 1–22.

[90] Srivaths Ravi, Anand Raghunathan, Paul Kocher, and Sunil Hattangady. 2004. Security in embedded systems: Design challenges. *ACM Transactions on Embedded Computing Systems (TECS)* 3, 3 (2004), 461–491.

[91] Steven K Reinhardt and Shubhendu S Mukherjee. 2000. Transient fault detection via simultaneous multithreading. In *Proceedings of the 27th annual international symposium on Computer architecture*. 25–36.

[92] Mário Zenha Rela, Henrique Madeira, and Joao Gabriel Silva. 1996. Experimental evaluation of the fail-silent behaviour in programs with consistency checks. In *Proceedings of Annual Symposium on Fault Tolerant Computing*. IEEE, 394–403.

[93] Abhishek Rhisheekesan, Reiley Jeyapaul, and Aviral Shrivastava. 2019. Control flow checking or not?(for soft errors). *ACM Transactions on Embedded Computing Systems (TECS)* 18, 1 (2019), 1–25.

[94] Mark Russinovich and Zary Segall. 1995. Fault-tolerance for off-the-shelf applications and hardware. In *Twenty-Fifth International Symposium on Fault-Tolerant Computing. Digest of Papers*. IEEE, 67–71.

[95] F Saglietti. 1990. Strategies for the Achievement and Assessment of Software Fault-Tolerance. *IFAC Proceedings Volumes* 23, 8 (1990), 303–308.

[96] Joshua Schiffman, Thomas Moyer, Hayawardh Vijayakumar, Trent Jaeger, and Patrick McDaniel. 2010. Seeding clouds with trust anchors. In *Proceedings of the 2010 ACM workshop on Cloud computing security workshop*. 43–46.

[97] R. Keith Scott, James W. Gault, and David F. McAllister. 1987. Fault-tolerant software reliability modeling. *IEEE transactions on Software Engineering* 5 (1987), 582–592.

[98] Aviral Shrivastava, Abhishek Rhisheekesan, Reiley Jeyapaul, and Carole-Jean Wu. 2014. Quantitative analysis of control flow checking mechanisms for soft errors. In *Proceedings of the 51st Annual Design Automation Conference*. 1–6.

[99] Mohammadreza Amel Solouki, Jacopo Sini, and Massimo Violante. 2023. An Experimental Evaluation of Control Flow Checking for Automotive Embedded Applications Compliant With ISO 26262. *IEEE Access* 11 (2023), 51185–51198. https://doi.org/10.1109/ACCESS.2023.3279731

[100] Wilfredo Torres-Pomales. 2000. Software fault tolerance: A tutorial. (2000).

[101] João P. Trovao. 2019. Trends in Automotive Electronics [Automotive Electronics]. *IEEE Vehicular Technology Magazine* 14, 4 (2019), 100–109. https://doi.org/10.1109/MVT.2019.2939757

[102] Jens Vankeirsbilck, Niels Penneman, Hans Hallez, and Jeroen Boydens. 2017. Random additive signature monitoring for control flow error detection. *IEEE transactions on Reliability* 66, 4 (2017), 1178–1192.

[103] Jens Vankeirsbilck, Niels Penneman, Hans Hallez, and Jeroen Boydens. 2017. Random additive signature monitoring for control flow error detection. *IEEE transactions on Reliability* 66, 4 (2017), 1178–1192.

[104] Jens Vankeirsbilck, Niels Penneman, Hans Hallez, and Jeroen Boydens. 2018. Random additive control flow error detection. In *International Conference on Computer Safety, Reliability, and Security*. Springer, 220–234.

[105] Raoul Velazco, Pascal Fouillat, and Ricardo Reis. 2007. *Radiation effects on embedded systems*. Springer Science & Business Media.

[106] Ramtilak Vemu and Jacob Abraham. 2011. CEDA: Control-flow error detection using assertions. *IEEE Trans. Comput.* 60, 9 (2011), 1233–1245.

[107] Rajesh Venkatasubramanian, John P Hayes, and Brian T Murray. 2003. Low-cost on-line fault detection using control flow assertions. In *9th IEEE On-Line Testing Symposium, 2003. IOLTS 2003*. IEEE, 137–143.

[108] Giuliana Santos Veronese, Miguel Correia, Alysson Neves Bessani, Lau Cheuk Lung, and Paulo Verissimo. 2011. Efficient byzantine fault-tolerance. *IEEE Trans. Comput.* 62, 1 (2011), 16–30.

[109] Massimo Violante, Cristina Meinhardt, Ricardo Reis, and Matteo Sonza Reorda. 2011. A low-cost solution for deploying processor cores in harsh environments. *IEEE Transactions on Industrial Electronics* 58, 7 (2011), 2617–2626.

[110] Massimo Violante, Cristina Meinhardt, Ricardo Reis, and Matteo Sonza Reorda. 2011. A Low-Cost Solution for Deploying Processor Cores in Harsh Environments. *IEEE Transactions on Industrial Electronics* 58, 7 (2011), 2617–2626. https://doi.org/10.1109/TIE.2011.2134054

[111] Jeffrey M Voas and Gary McGraw. 1997. *Software fault injection: inoculating programs against errors*. John Wiley & Sons, Inc.

[112] Fan Wang and Vishwani D. Agrawal. 2008. Single Event Upset: An Embedded Tutorial. In *21st International Conference on VLSI Design (VLSID 2008)*. 429–434. https://doi.org/10.1109/VLSI.2008.28

[113] Torres Wilfredo. 2000. Software fault tolerance: A tutorial. (2000).

[114] Ying C Yeh. 1996. Triple-triple redundant 777 primary flight computer. In *1996 IEEE Aerospace Applications Conference. Proceedings*, Vol. 1. IEEE, 293–307.

[115] Ze Zhang, Sunghyun Park, and Scott Mahlke. 2020. Path sensitive signatures for control flow error detection. In *The 21st ACM SIGPLAN/SIGBED Conference on Languages, Compilers, and Tools for Embedded Systems*. 62–73.

[116] Bowen Zheng, Yue Gao, Qi Zhu, and Sandeep Gupta. 2015. Analysis and optimization of soft error tolerance strategies for real-time systems. In *2015 International Conference on Hardware/Software Codesign and System Synthesis (CODES+ISSS)*. IEEE, 55–64.
[117] Zhiqi Zhu, Joseph Callenes-Sloan, and Benjamin Carrion Schafer. 2018. Control Flow Checking Optimization Based on Regular Patterns Analysis. In *2018 IEEE 23rd Pacific Rim International Symposium on Dependable Computing (PRDC)*. IEEE, 203–212.
[118] Haissam Ziade, Rafic A Ayoubi, Raoul Velazco, et al. 2004. A survey on fault injection techniques. *Int. Arab J. Inf. Technol.* 1, 2 (2004), 171–186.